\documentclass[twocolumn,prb]{revtex4}
\usepackage{epsfig}
\usepackage{amsmath}
\usepackage{amssymb}
\usepackage{graphicx}
\usepackage[english]{babel}
\begin{document}
\title{Modelling Thickness-Dependence of Ferroelectric Thin Film Properties}
\author{L. P\'alov\'a}
\author{P. Chandra}
\author{K.M. Rabe}
\affiliation{Center for Materials Theory, Department of Physics and Astronomy, Rutgers University, 
Piscataway, NJ 08854}
%
\begin{abstract}
We present a segregated strain model that can describe the thickness-dependent
dielectric properties of epitaxial ferroelectric films.  Using a 
phenomenological Landau approach, we present results for two specific 
materials, making comparison with experiment and with 
first-principles calculations whenever possible.  We also 
suggest a ``smoking gun'' benchtop probe 
to test our model.
\end{abstract}

\maketitle

\section{Introduction}

Increasing demands for high-density data storage with ultra-fast 
accessibility present tremendous challenges.  In parallel to the 
characterization of new materials, important size-dependent 
effects must be understood to optimize design.  This is 
particularly true for ferroelectric memories, 
whose nonvolatility and low power consumption make them well-suited 
for portable applications;~\cite{Auciello98,Scott00} 
their dielectric properties are strongly 
dependent on electromechanical boundary conditions due to the 
long-range nature of their underlying electrostatic interactions.
The sensitivity of ferroelectricity to 
homogeneous strain in bulk perovskite oxides is 
well-known.~\cite{Lines77}                      
In thin films, the effects of homogeneous misfit strain
have been identified,\cite{Pertsev98} studied and controlled to the 
point that particular 
systems have been strain-engineered to have spontaneous polarizations 
significantly larger than those in the bulk.~\cite{Choi04,Haeni04}  
Despite these impressive
achievements, several authors~\cite{Balzar04,Alpay04,Catalan04,Bratkovsky05} have suggested that 
homogenous epitaxial strain
cannot qualitatively account for all the observed effects in ferroelectric
films.  In particular, the thickness-dependence of their dielectric 
properties has been attributed to defect-induced strain 
gradients.~\cite{Balzar04,Catalan04}
In this paper we explore whether these observed size-effects
are also consistent with an alternative scenario
where the majority of each ferroelectric film is 
homogeneously strained.
Our phenomenological study indicates that thickness-dependent
dielectric measurements are insufficient to determine the
presence/absence of underlying inhomogeneous strain and 
we suggest further benchtop experiments that will resolve this issue.

In well-screened coherent epitaxial ferroelectric thin films, 
uniform polarization
is energetically favored. Lattice mismatch between 
the film and the
substrate is a key source of macroscopic strain in these 
systems.~\cite{Pertsev98,Dawber05}
For very thin films, the energy cost for homogeneously
straining the film to match the substrate is less than that associated
with the introduction of stress-relieving defect structures at the interface.
However in thicker films, such misfit dislocations form and produce
strain gradients~\cite{Balzar04,Dawber05,Suzuki99,Sun04,Nagarajan05};
threading dislocations and point defects are additional
sources of inhomogeneous strain.  
In planar ferroelectric films, 
inhomogeneous strain can affect the ferroelectric transition 
through both smearing and through its coupling to the polarization. 
Such flexoelectric coupling
of strain gradients to the polarization has been the topic of
much recent interest~\cite{Catalan04,Ma01,Gruverman03}
particularly as it has been suggested that
flexoelectric effects are enhanced by large dielectric 
coefficients.~\cite{Tagantsev85}  Recently it has been argued that
such strain/stress gradients are crucial for the modelling of
thickness-dependent dielectric properties of
ferroelectric films.~\cite{Balzar04,Catalan04}  
Here we propose an alternative model: that the misfit dislocations
reside within a thin buffer layer next to the interface; the 
majority of the film,
which is relatively defect-free, is then
homogeneously strained.~\cite{Speck94} In our phenomenological treatment, we
also include a bulk anisotropy~\cite{Bratkovsky05} in the form 
of an effective field,~\cite{Vendik00} 
possibly due to asymmetry
of the electrodes and/or to the thin buffer layer. 
We model the thickness-dependent dielectric
properties in two different types of ferroelectric films,
and compare our results with experiment and with first-principles
calculations whenever possible.  Finally we discuss a benchtop ``smoking
gun'' probe to distinguish our segregated strain scenario with that of 
inhomogeneous strain in ferroelectric thin films.

The structure of this paper is as follows.  In Section II we review 
the experiments that motivate this study and their implications
for any descriptive model. Details of
our phenomenological Landau approach are presented in Section III,
with specific discussion of the appropriate boundary conditions
and depolarization effects.  In Section IV we present our main results
for films of two specific materials, with comparison to previous
findings whenever possible and predictions for future measurements.
The implications of our model and our results 
are discussed 
in Section V. We end with a summary (Section VI) and with ideas for
future work.

\section{Experimental Motivation}
\label{Sec II}

Broadening of the temperature-dependent permittivity 
in thin films near the paraelectric-ferroelectric transition
is reported by several experimental
groups;~\cite {Ahn97,Shaw99,Paruch01,Sinnamon02,Lookman04}
this observed smearing, accompanied by an overall reduction in
its magnitude, is more pronounced with decreasing film 
thickness.  
Careful measurements on free-standing ferroelectric lamellae
yield bulk-type dielectric responses, suggesting
interfacial effects as the source of these thickness-dependent
effects.~\cite{Saad04}  
A second related observation is that
there is a clear separation of temperature scales associated with
the onset of reversible spontaneous polarization and the maximum
of the dielectric constant in thin ferroelectric films.~\cite{Catalan04}

\begin{figure} [t!]
\begin{center}
\includegraphics[scale=0.5]{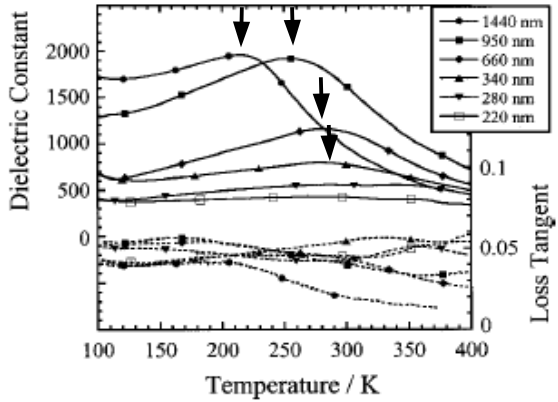}
\caption[parkergregg]
{Relative permittivity (dielectric constant) as a function of temperature
($\epsilon (T)$) for $(Ba,Sr)TiO_3$ (BST) films 
of different thicknesses 
grown on $SrRuO_3/MgO$ (SRO) substrates~\cite{Sinnamon02} where 
the $\epsilon_{max} (T)$ at temperature $T_{max}$ is indicated with an arrow.
}
\label{pgfig}
\end{center}
\end{figure}

In Fig.~\ref{pgfig} we display relative permittivity measurements
done on $(Ba,Sr)TiO_3$ (BST) 
thin films grown 
on $SrRuO_3$ (SRO).~\cite{Sinnamon02}
The measurements show suppression of the relative permittivity 
with decreasing film thickness.
As the film thickness decreases to 340 nm,
the temperature associated with the maximum of the permittivity ($T_{max}$)
appears to saturate at about 300 K,
with $T_{max}$ for the two thinnest films not being clearly discernable.
As we will discuss in Section IV, this is consistent with the prediction of our model that there should be a maximum in $T_{max}(l)$ at a thickness $l=l^*$; from the data in Figure \ref{pgfig} we estimate $l^* < 340$ nm in BST films grown on SRO.

In ferroelectric films, in contrast with their bulk counterparts, 
there is an observed
distinction~\cite{Catalan04}
between $T_{max}$ and $T_{ferro}$, 
the temperature where polarization becomes switchable. 
This separation of temperature-scales and the 
permittivity broadening discussed earlier are both features 
characteristic of dielectric behavior in
an applied bias field~\cite{Vendik00}; the latter could be due to
a real charge distribution or it could result from
another physical effect~\cite{Bratkovsky05} that breaks the 
symmetry $P \rightarrow - P$.
For example it has been noted~\cite{Catalan04} 
that flexoelectric coupling, known to increase
near a ferroelectric transition,~\cite{Ma01}
implies a spatially-varying effective
field term due to the underlying inhomogeneous strain~\cite{Catalan04}.
The resulting phenomenological model successfully reproduces 
key thickness-dependences of the dielectric properties.~\cite{Catalan04}
Here we ask whether these experimental trends are indeed proof of
underlying strain inhomogeneities, or whether they may be consistent
with another strain profile.

\section{Landau Theory}

\begin{figure}
\includegraphics[scale=0.29]{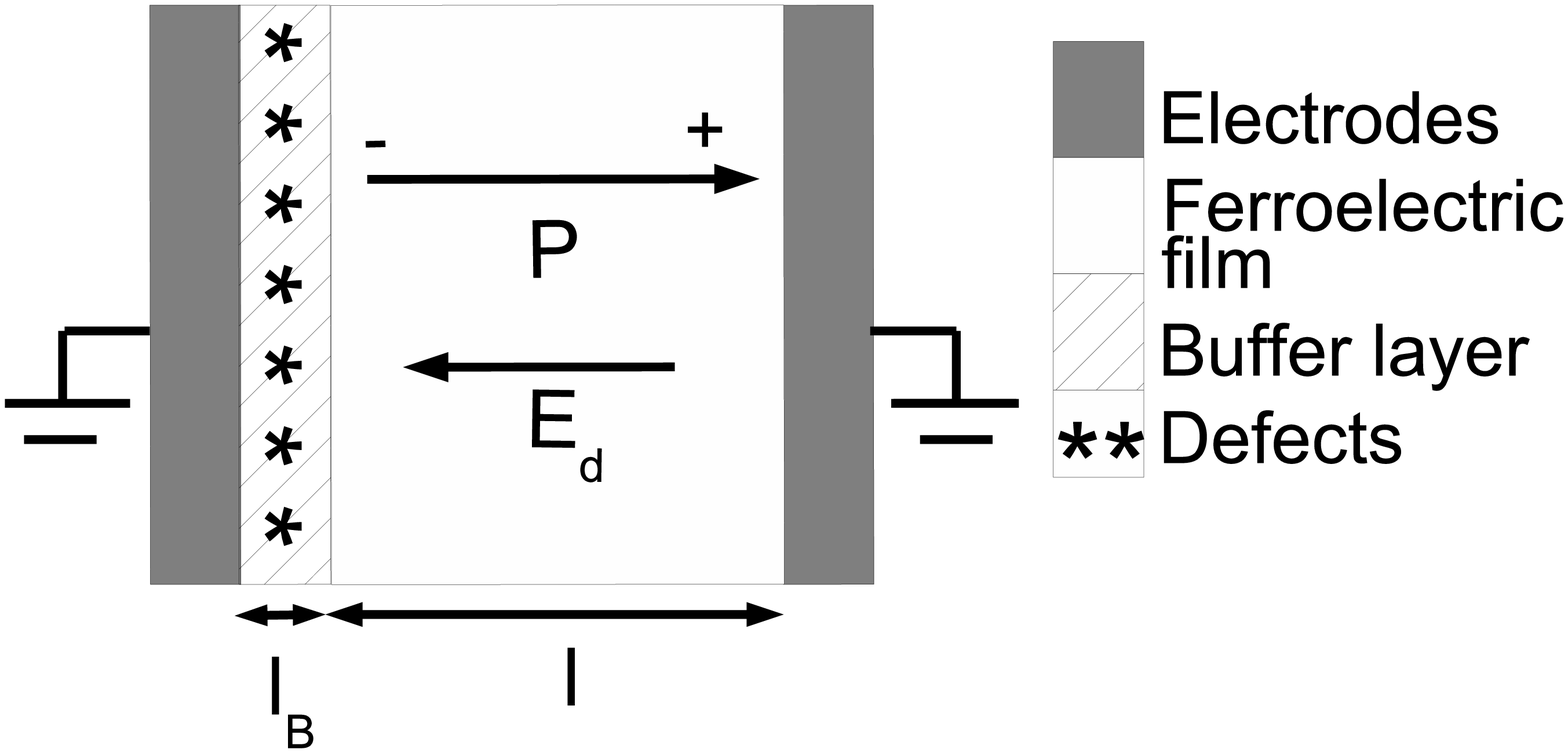}
\caption[ert]
        {A schematic of the planar ferroelectric capacitor under consideration,
with the key length-scales and regions clearly demarcated.  Note that
the mismatch defects are segregated in a buffer layer of thickness $l_B$
and that the polarization and strain are homogeneous
in the majority of the film. Incomplete charge compensation at the
ferroelectric-electrode interfaces results in a residual depolarization field, as shown.}
\label{filmfig}
\includegraphics[scale=0.33]{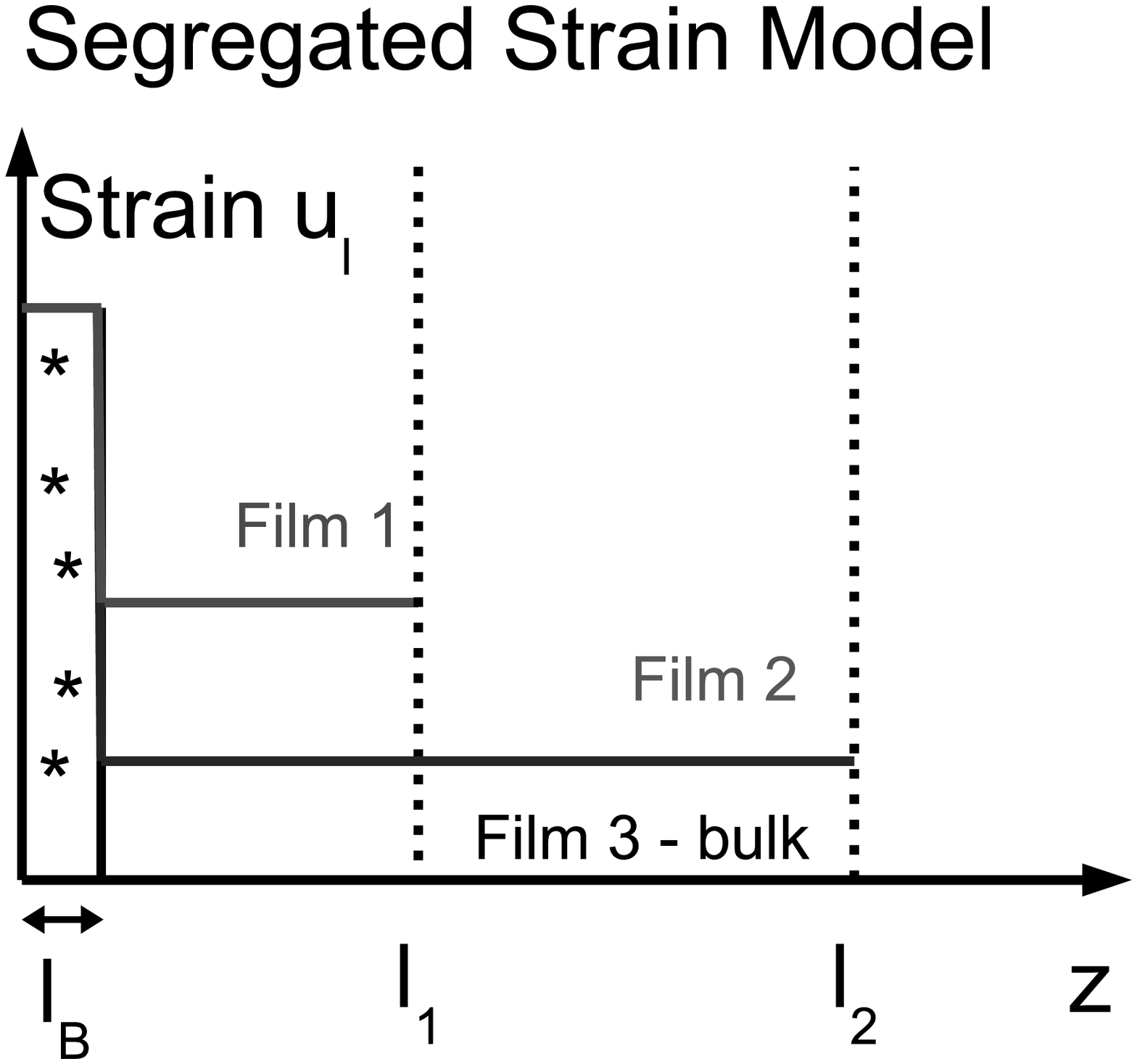}
\caption[homogstrain2]
{Schematic of the segregated strain model; 
here the 
elastic defects
reside within $l_B$ ($<< l$)  of
the film-substrate
interface so that the majority of
the film is homogeneously strained.}
\label{homogstrain2fig}
\end{figure}

We model the thickness-dependent properties of strained
ferroelectric films using a Landau
approach~\cite{Lines77,Chandra06} where 
all misfit dislocations are assumed to reside within 
a thin buffer layer of size $l_B << l$ where $l$ is
the film thickness (e.g. Fig.~\ref{filmfig});
this is in contrast to other approaches where
these defects are assumed to be roughly 
uniformly distributed within the film.~\cite{Balzar04}
Elastic relaxation then occurs so that there is homogeneous strain 
in the film except for
the buffer layer (e.g. Fig.~\ref{homogstrain2fig}). 
Recent X-ray diffraction experiments~\cite{Sinnamon02} support 
previous
suggestions~\cite{Kim99} that the in-plane film strain, $u_l$,
decreases with increasing thickness ($l$) of the overall films; 
furthermore the observed thickness-dependent strain fits an exponential 
form~\cite{Kim99,Sinnamon02} very well.  Therefore  
in our segregated strain approach, thin films 
experience homogeneous film strain that decreases
exponentially with their overall thickness $l$;
more specifically we model this thickness-dependent in-plane
film strain as
%
%
\begin{equation}
\label{eqnul}
u_l = u_m e^{-l/l_c}, \quad u_m = (b - a)/b,
\end{equation}
where $l_c$ is a characteristic length-scale of the strain relaxation,
and $a$ and $b$ refer to the in-plane lattice parameters of the
film (bulk value) and the substrate, respectively~\cite{Speck94}.
We emphasize that our values of $l_c$ are determined from X-ray
diffraction~\cite{Sinnamon02,Kim99} for films of a range
of thicknesses.  We note that these values of $l_c$ are comparable
to the film thicknesses of interest ($l_c \sim l$), resulting in a
non-trivial thickness-dependence of the strain-related properties.
More generally, we remark that the segregated strain approach described
here has been 
used in modelling epitaxially strained superlattices~\cite{Johnston05}
with results that are consistent with experiment.~\cite{Rios03}

In Fig.~\ref{filmfig} we display a schematic of the planar ferroelectric
capacitor under consideration with the length-scales involved.
More specifically we assume a 
single-domain ferroelectric film with the (uniaxial) polarization in the 
z-direction, normal to the film-substrate interface.
Physically we expect a build-up of free surface charge at the ferroelectric
boundaries which, if uncompensated, results in a depolarization field 
$E_d$. In practice such depolarization
effects are significantly reduced by metal electrodes that provide charge
compensation; however their non-ideal nature means that some residual
depolarization field remains. 
Because $E_d$ is proportional to $\frac{l_e}{l}$, where $l_e$ is
the screening length of the electrodes, 
its importance increases with
reduced film thickness~\cite{Dawber05,Chandra06,Mehta73,Kretchmer79}
and becomes significant~\cite{Dawber03,Junquera03} for $l \sim 100 \rm{nm}$;
a term in our Landau expansion will be included to account
for these depolarization effects.

The defect buffer layer is also displayed in Fig.~\ref{filmfig} and,
as discussed above, we assume homogeneous (but possibly relaxed relative to the substrate)
strain in the remainder of the ferroelectric film which is controlled
by the substrate-film lattice mismatch.  
Following a previous analysis of epitaxially 
strained films,~\cite{Pertsev98}
the stress tensors $\sigma_{zz}=\sigma_{xz}=\sigma_{yz}=0$, 
because there are no tractions acting on the top film surface.
For the special case of a (001) ferroelectric film grown on a cubic
substrate,
the strain $u_{xx} = u_{yy} = u_l$ and $u_{xy}=0$
since the angle between the two lattice vectors remains 
unchanged ($\theta = \frac{\pi}{2})$.
We consider here
film tetragonal symmetry with $u_{zz} \neq 0$ and $u_{yz} = u_{xz} = 0$.
These mixed mechanical boundary conditions associated with two-dimensional
clamping mean that the standard elastic free energy $G (P,\sigma)$
cannot be used to find the equilibrium properties of these systems; instead
a Legendre transformation, 
$G'(P,u_l) = G(P,\sigma) + u_l(\sigma_{xx}+\sigma_{yy}) + u_{zz}\sigma_{zz}$,  
to a modified thermodynamic potential must be performed in order to study
the equilibrium properties of the constrained film.~\cite{Pertsev98}

We are almost ready to write an expression for $G'(P,u_l)$ and to calculate
observable thermodynamic quantities.  As discussed earlier, 
the experiments suggest a term
in $G$ that breaks the 
symmetry $P \to -P$; 
this is achieved by linear 
coupling of $P$ to an external electric
field $E_{ext}$ and/or to an effective bias field~\cite{Bratkovsky05}
which we take to be of the form
\begin{equation}
\label{eqnWl}
W_l=W_0 e^{-l/l_w},
\end{equation}
where 
$l_w \sim l_c$.  We note that the thickness-dependence
of $W_l$ is included to model the increased smearing of
the dielectric susceptibility with decreasing $l$
of ferroelectric films.~\cite{Bratkovsky05}
At present we will treat $W_l$ phenomenologically, and will
defer discussion of its exponential decay and its possible origins
to Section~\ref{discussionsec}.

Putting all these elements together, 
we begin our phenomenological study with 
the free-energy expansion 
\begin{eqnarray} 
\label{eqnG'}
 G(P,\bar\sigma,T) & = & \frac{1}{2} \tilde{\alpha} (T) P^2 +\frac{1}{4} \gamma P^4 - (W_l +E_{ext})P \nonumber \\ 
 & & - Q_{11} \sigma_{zz} P^2 - Q_{12} (\sigma_{xx} + \sigma_{yy}) P^2 \nonumber \\
 & & - \frac{1}{2}s_{11} (\sigma_{xx}^2 + \sigma_{yy}^2 + \sigma_{zz}^2)
- s_{12} (\sigma_{xx}\sigma_{yy} ) \nonumber \\
 & & - s_{12} \sigma_{zz} (\sigma_{xx} + \sigma_{yy}) - \frac{1}{2} s_{44} \sigma_{xy}^2, 
\end{eqnarray}
where $\tilde{\alpha} (T) = \alpha (T) + \alpha_d$; $\alpha (T) = \beta (T-T_{bulk})$, $T_{bulk}$ is the bulk transition temperature, $\alpha_d$ is discussed below, and
$\beta$ and $\gamma$ are 
Landau coefficients;
here $Q_{ij}$ and $s_{ij}$ are the electrostrictive constants
and the elastic compliances at constant polarization respectively.
The depolarization field contributes to the free energy through the coefficient $\tilde{\alpha} (T)$ in Eq.(\ref{eqnG'})~\cite{Chandra06,Tilley92}
\begin{equation}
\label{eqnalphad}
\alpha_d = \frac{l_e}{\epsilon_0 \epsilon_e l},
\end{equation}
where $l_e$ is the screening length of the electrodes, and 
$\epsilon_0$ and $\epsilon_e$ are the electric permittivities of the vacuum 
and the electrodes respectively.

The mechanical conditions in the film are 
$\partial G/\partial \sigma_{xx} = \partial G/\partial \sigma_{yy} = -u_l$,
$\partial G/\partial \sigma_{xy}=0$ and
$\partial G/\partial \sigma_{zz}=-u_{zz}$~\cite{Pertsev98}.
Solving for the in-plane stresses, one finds that 
$\sigma_{xy} = 0$ and
$\sigma_{xx} = \sigma_{yy} \equiv \sigma$,
where the applied in-plane
stress $\sigma$ is eliminated by the in-plane strain $u_l(\sigma)$. 
This procedure, together with 
$\sigma_{zz} = 0$,  
leads to
\begin{eqnarray} 
\label{freeenergy}
G'(P_l,u_l,T) & = & \frac{u^2_l}{s_{11}+s_{12}}
+ \frac{1}{2} \alpha^*_l(T) P_l^2 \nonumber \\
 & & +\frac{1}{4} \gamma^* P_l^4 - (W_l + E_{ext})P_l,
\end{eqnarray}
where $\gamma^* = \gamma + \frac{4 Q_{12}^2}{s_{11}+s_{12}}$ and
$\alpha^*_l(T) = \alpha (T) - u_l\frac{4 Q_{12}}{s_{11}+s_{12}} + \alpha_d$.
We note that we explicitly
refer to the $l$-dependence of the polarization ($P_l$), which here
results from the thickness-dependence of the strain ($u_l$), 
the bias field ($W_l$) and depolarization field term ($\alpha_d$),
consistent with observation.~\cite{Catalan04}
One can express the out-of-plane strain ($u_{zz}$) through its dependence 
on the out-of-plane polarization ($P_l$) and the in-plane strain ($u_l$) as 
\begin{equation}
\label{eqnuzz}
u_{zz} (P_l,u_l) = \frac{2 s_{12}}{s_{11} + s_{12}} u_l + 
\big( Q_{11} - \frac{2 s_{12}}{s_{11}+s_{12}} Q_{12} \big) P_l^2.
\end{equation}

The Curie film temperature $T_c^*$ refers to the paraelectric-ferroelectric 
transition 
at zero total field, 
$E_l^T \equiv W_l+E_{ext} = 0$.
It increases with an applied compressive strain ($Q_{12} < 0$) 
\begin{equation} 
\label{eqnTc}
T_c^* = T_{bulk} + \frac{1}{\beta} \Big [
u_l\frac{4 Q_{12}}{s_{11}+s_{12}} - \frac{l_e}{\epsilon_0 \epsilon_e l}\Big]
\end{equation}
but has a decreasing component for very thin films due to depolarization
effects. 
The dielectric susceptibility is 
\begin{equation} \label{eqnchi}
\chi^{-1}_l = \epsilon_0 \frac{d^2G}{dP_l^2} = \epsilon_0 \Big[\alpha^*_l(T) + 3\gamma^* P_l^2 \Big]
\end{equation}
which diverges at $T_c^*$ if the spontaneous polarization $P_l \rightarrow 0$;
that can only occur if $E_l^T=0$ (see Eq.~\ref{eqnP} below).
The dielectric susceptibility is observed to diverge for 
bulk systems~\cite{Lines77} 
and for free-standing films.~\cite{Saad04}
In general 
$E_l^T \neq 0$ for ferroelectric capacitors and $P_l \neq 0$ at $T_c^*$,
so that $\chi_l$  
has a finite maximum at a temperature $T_{max}$ defined by
$\frac{\partial \chi_l}{\partial T} \vert_{T=T_{max}} = 0$.
We note that this condition combined with the expression for $\chi_l$
above yields 
\begin{equation}
\label{eqnTmax}
P_l(T) \frac{dP_l(T)}{dT}\Big|_{T_{max}} = -\frac{\beta}{6 \gamma^*}
\end{equation}
where we see that $T^l_{max}$ differs from $T_c^*$ and
depends on film thickness
via the polarization;
this equation generally has to be solved numerically to obtain $T_{max}^l$ once
the expression for $P_l (T)$ has been determined.

The condition for finding the system in its equilibrium state is 
$\frac{\partial G}{\partial P}=0$.
The spontaneous polarization $P_l$ emerges then as the solution(s) 
to the following 
cubic equation
\begin{equation} \label{eqnP}
\alpha^*_l(T) P_l + \gamma^* P_l^3 = E_l^T
\end{equation}
where for $E_l^T =W_l + E_{ext} \neq 0$, we have to be careful 
to distinguish between the
paraelectric (nonswitchable) polarization 
$P_p$
and the ferroelectric
(switchable) polarization 
$P_f$;
here switchable refers
to the fact that there are multiple solutions for the polarization
that can be accessed by application of a finite $E_{ext}$.
There are three solutions to the equation (\ref{eqnP})
\begin{equation} \label{eqnPpara}
P_p = \Big(\frac{E_l^T}{2\gamma^*} + \sqrt{\cal R}\Big)^{1/3} - 
\Big(-\frac{E_l^T}{2\gamma^*} + \sqrt{\cal R}\Big)^{1/3}
\end{equation}
and
\begin{equation} \label{eqnPferro}
P_f =\frac{1}{2}P_p \pm i \frac{\sqrt{3}}{3} \Bigg(
\Big(\frac{E_l^T}{2\gamma^*} + \sqrt{\cal R}\Big)^{1/3} + \Big(-\frac{E_l^T}
{2\gamma^*} + 
\sqrt{\cal R}\Big)^{1/3}\Bigg),
\end{equation}
where
\begin{equation} \label{root}
{\cal R}  \equiv \frac{\alpha^{*3}_l(T)}{27\gamma^{*3}} + \frac{(E_l^T)^2}{4\gamma^{*2}}
\end{equation}
and the number of polarization solutions is determined by the sign
of ${\cal R}$ so that the single nonswitchable
$P_p$ corresponds to ${\cal R} > 0$.
Therefore the transition temperature $T_{ferro}$ between
nonswitchable and switchable polarization occurs when
${\cal R} = 0$ leading to the expression
\begin{equation} \label{eqnTferro}
T_{ferro} = T_c^* - \frac{3}{\beta}\Big(\frac{\gamma^*}{4}\Big)^{1/3}(E_l^T)^{2/3}.
\end{equation}
At this temperature, the paraelectric solution becomes an unstable extremum.

In general, the three temperature-scales
$T_{ferro}$, $T_{max}$ and $T_c^*$ differ as indicated in 
Fig.~\ref{templfig}.  We note that for 
very thin films
($<60$ nm), there is suppression of all three temperatures 
due to depolarization effects.  We also remark on the presence
of a maximum in $T_{max} (l)$ that has already been alluded
to in Section II; this feature will be discussed in more detail when
we apply this phenomenology to specific materials and substrates.

\section{Results}
\label{Resultssec}

\begin{table}[!t]
\begin{center}
\caption{Landau parameters for $BST$~\cite{Pertsev98,Catalan04} and 
$STO$~\cite{Chen06} (in SI units). We use the $T > 100 K$ values for 
$\alpha(T)$, except in calculations in Fig.~\ref{tferrolstrainstofig} where 
we interpolate the $T >100 K$ and the $T < 50 K$ values
to $75 K$ where the two $\alpha(T)$ functions cross.} 
\label{LGtab}
\begin{tabular}{|c|c|c|c|c|}
\hline
 film & $\gamma(T)$ [$10^6$] & $Q_{11}$ & $Q_{12}$ & 
$s_{11} + s_{12}$ [$10^{-12}$]\\
\hline
 BST & $4(796 + 2.16(T-273))$ & 0.110 & $-0.0430$ & $5.6$\\
 STO & $1700$ & 0.066 & $-0.0135$ & $3.0$ \\
\hline
\end{tabular}
\begin{tabular}{|c|c|}
\hline
 film & $\alpha(T)$ [$10^5$] \\
\hline
 BST & $9.1 (T - 235.0)$ \\
 STO ($T > 100 K$) & $7.06 (T - 35.5)$ \\
 STO ($T < 50 K$) & $263.5$ (Coth$[42.0/T] - 0.90476$) \\
\hline
\end{tabular}
\caption{Film parameters: effective field $W_0$  
and compressive strain $u_m$~\cite{Haeni04,Catalan04} with associated length-scales 
$l_w$ and $l_c$~\cite{Sinnamon02} (see Eqs.~\ref{eqnul}
 and~\ref{eqnWl}); also the values for screening length $l_e$~\cite{Lichtensteiger05}
 and the relative permittivity $\epsilon_e$ of electrodes are shown.} 
\label{Wumtab}
\begin{tabular}{|c|c|c|c|c|c|c|c|}
\hline
 film & substrate & $W_0$ [$kV/cm$] & $l_w$ [nm] & $u_m$ [$\%$] & $l_c$ [nm] & $l_e$ [nm] & $\epsilon_e$\\
\hline
 BST & SRO & $400$ & $300$ & $-0.50$ & $300$ & $0.023$ & $1.0$\\
 BST & PSS & $450$ & $300$ & $-0.77$ & $300$ & $0.400$ & $1.0$\\
 STO & LSAT & $400$ & $300$ & $-0.90$ & $300$ & $0.023$ & $1.0$\\
\hline
\end{tabular}
\end{center}
\end{table}

In this section, we calculate dielectric properties for two
specific materials, $(Ba_{0.5}Sr_{0.5})TiO_3$ (BST) and $SrTiO_3$ (STO).
Our study of BST films  
allows us to make direct comparison between our calculated properties
and the experiments (Fig.~\ref{pgfig})
that motivated the inhomogeneous strain scenario.~\cite{Catalan04}
In order to explore different parameter regimes, we study these films
on two distinct substrates, $SrRuO_3$ (SRO) and $Pt/SiO_2/Si$ (PSS);
here we note that the latter is a hypothetical case since to date
epitaxially-grown single-crystal films of BST on PSS have not yet been 
realized.

We also apply our phenomenological treatment to STO films 
that are known for their coherence; this is achieved by highly controlled
growth conditions that inhibit defect 
formation and thus inhomogeneous strain effects 
are not expected.~\cite{Haeni04}
To our knowledge, 
there do not exist 
published 
high-resolution 
dielectric measurements of strained STO films
with polarization normal to the electrode-ferroelectric interface.
We therefore compare our results whenever possible 
to first-principles calculations,~\cite{Antons05} and make experimental
predictions for a range of epitaxial strain values that could be realized
by a variety of substrates.

The parameters used in our calculations are presented 
in Tables~\ref{LGtab} and~\ref{Wumtab}.
Table~\ref{LGtab} indicates the Landau coefficients used for each material. 
Film-related parameters, displayed in Table~\ref{Wumtab},
are determined from data on strain relaxation; the characteristic length 
$l_c$ (see Eq.~\ref{eqnul}) from the lattice constant measurements lies 
somewhere between $200$ and $300$ nm~\cite{Sinnamon02,Kim99}. In order
 to make comparison with the inhomogeneous strain model scenario, which uses
 a characteristic length scale $300$ nm~\cite{Catalan04}, we keep this $l_c$
value in our calculations.
Data on electrode screening lengths ($l_e$) is $l_e$ of 
$SrRuO_3$ (SRO) $0.23$ \AA~\cite{Lichtensteiger05}, and
we set $l_e$ of 
$(LaAlO_3)_{0.29} \times (SrAl_{0.5}Ta_{0.5}O_3)_{0.71}$ (LSAT) to be the
same 
value, since LSAT is dominated by $SrAl_{0.5}Ta_{0.5}O_3$ that
is very similar to SRO.
We choose the screening length of $Pt/SiO_2/Si$ (PSS) to be 
$l_e = 4$ \AA, which
is expected to be larger than $l_e$ in metallic SRO due to the presence of 
semiconducting silicon. 
We display strain $u_m$ value for
three 
different substrates, BST on SRO,~\cite{Catalan04}
BST on PSS ($a_{BST}=3.95$ \AA, $b_{Pt}=3.92$ \AA) 
and STO on LSAT.~\cite{Haeni04}
The bias field is set to be 
$W_0 = 4.0\times 10^7$ V/m for BST films on SRO, a value that is 
comparable to that of applied external fields in related BST 
dielectric measurements.~\cite{Sinnamon02} We keep the same $W_0$
for STO films
and a slightly different one ($W_0 = 4.5\times 10^7$ V/m) 
for BST films on PSS.
The bias field $W_l$ is treated phenomenologically (see Eq.~\ref{eqnWl}) and 
we emphasize its crucial role in modelling key features of the
dielectric properties of ferroelectric films as will be discussed in 
more detail shortly.

\subsection{BST}
\label{BSTsec}

\begin{figure} [t!]
\begin{center}
\includegraphics[angle=-90,scale=0.33]{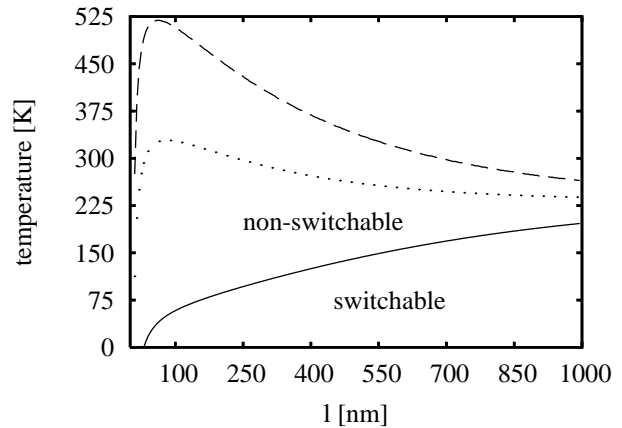}
\caption[templ]
{Thickness-dependence of the three distinct temperature-scales 
$T_{ferro}$ (solid), 
$T_{max}$ (dashed) and 
$T_c^*$  
(dotted line) with $E_{ext} = 0$ in the segregated defect model
described in the text; here Landau coefficients for
BST on SRO (see Tables~\ref{LGtab} and~\ref{Wumtab} in Section IV) have been used 
and $T_{max}(l)$ is noted to display a peak at $l^*$ = 60 nm.}
\label{templfig}
\end{center}
\end{figure}

\begin{figure} [t!]
\includegraphics[angle=-90,scale=0.33]{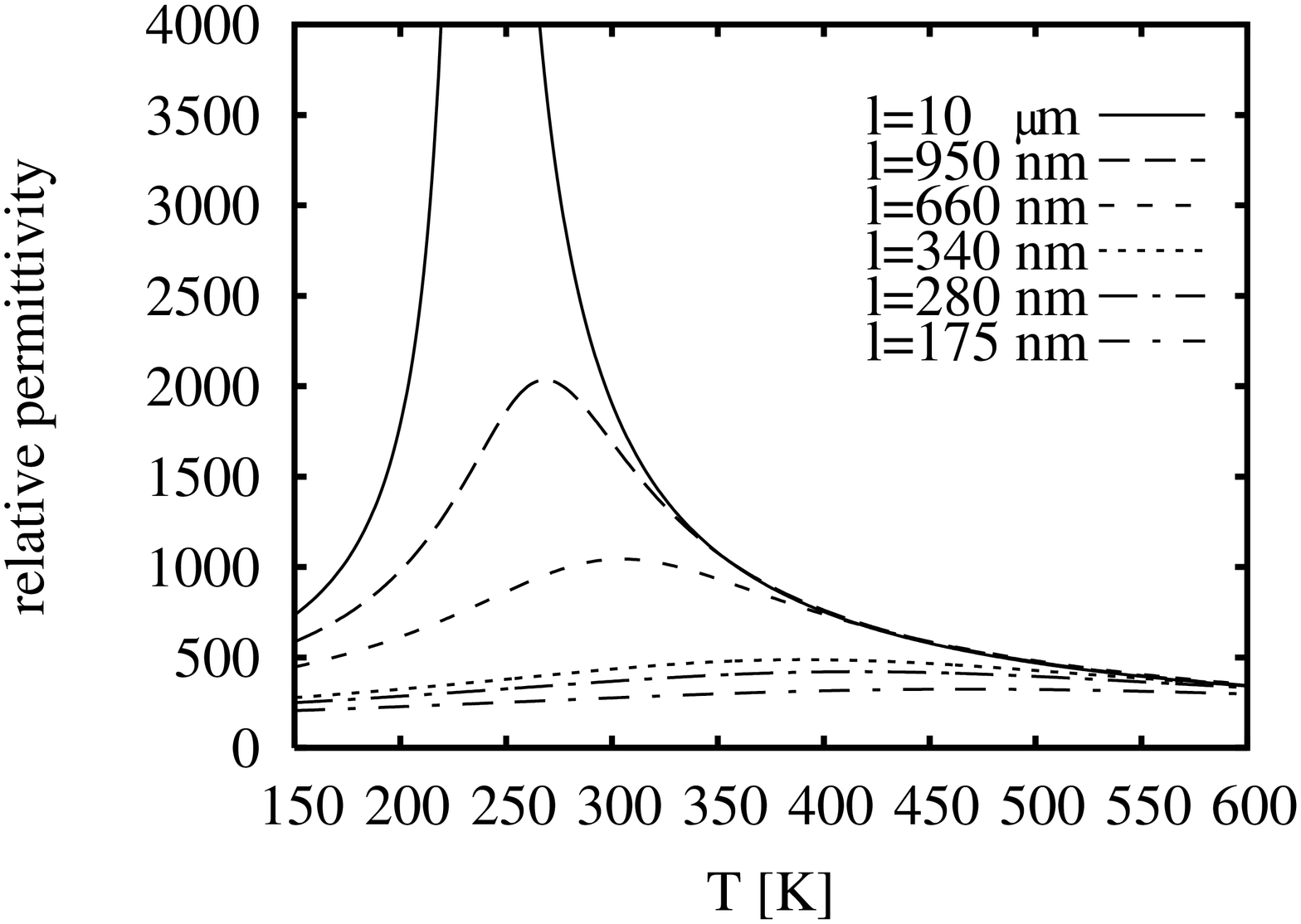}
\caption[ert]
        {Calculated relative permittivity as a function of temperature for
 $BST$ films on $SRO$ substrates of thicknesses 
$l=10\mu m$, $950$, $660$, $340$, $280$ and $175$ nm with $E_{ext}=0$.
The highest permittivity corresponds to the thickest film; the
divergence for $10\mu$m film at the bulk Curie temperature $235$ K
is indicated. 
Reduction
in the permittivity for thin films is observed;
the peak of the permittivity shifts towards higher temperatures in agreement with 
Fig.~\ref{pgfig}.}
\label{ertbstfig}
\includegraphics[angle=-90,scale=0.33]{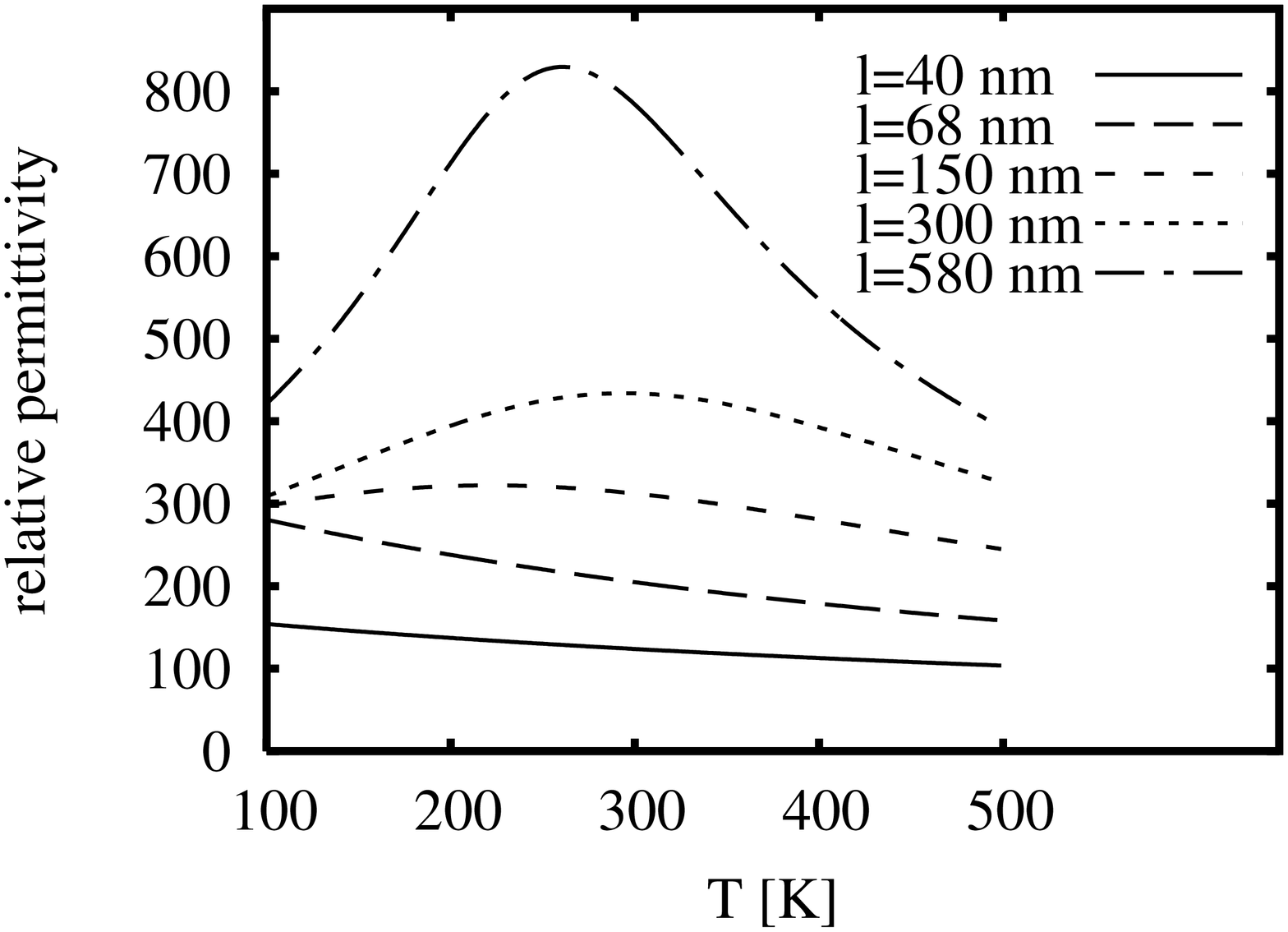}
\caption[ertpt]
        {Calculated relative permittivity as a function of temperature for
 $BST$ films epitaxially grown on PSS 
substrates of thicknesses 
$l=40$, $68$, $150$, $300$ and $580$ nm with $E_{ext}=0$.
The highest permittivity corresponds to the thickest film. Reduction
in the permittivity for thin films is observed, and
the peak of permittivity shifts towards lower temperatures.
}
\label{ertbstptfig}
\end{figure}

\begin{figure} [t!]
\includegraphics[angle=-90,scale=0.33]{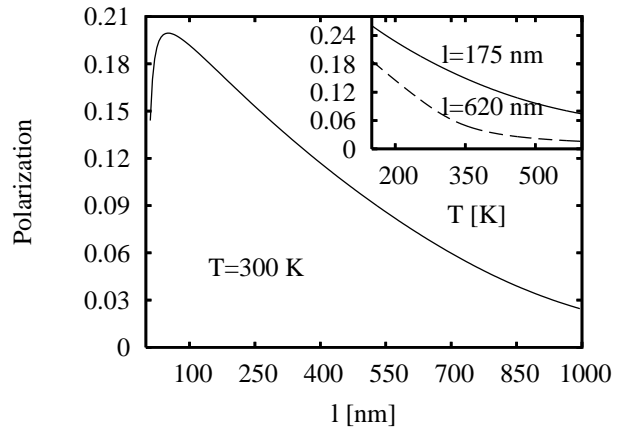}
\caption[pl300apt]
        {Calculated nonswitchable polarization P[$\frac{C}{m^2}$]
(see Fig.~\ref{templfig}) for $BST$ films on $SRO$ substrates of
various thicknesses ($l$) at temperature 300 K and with $E_{ext}=0$;
polarization as a function of temperature for films of 
thickness 175 (solid) and 620 nm (dashed line) are shown in the inset.}
\label{pl300aptbstfig}
\end{figure}

\begin{figure}[t!]
\includegraphics[angle=-90,scale=0.33]{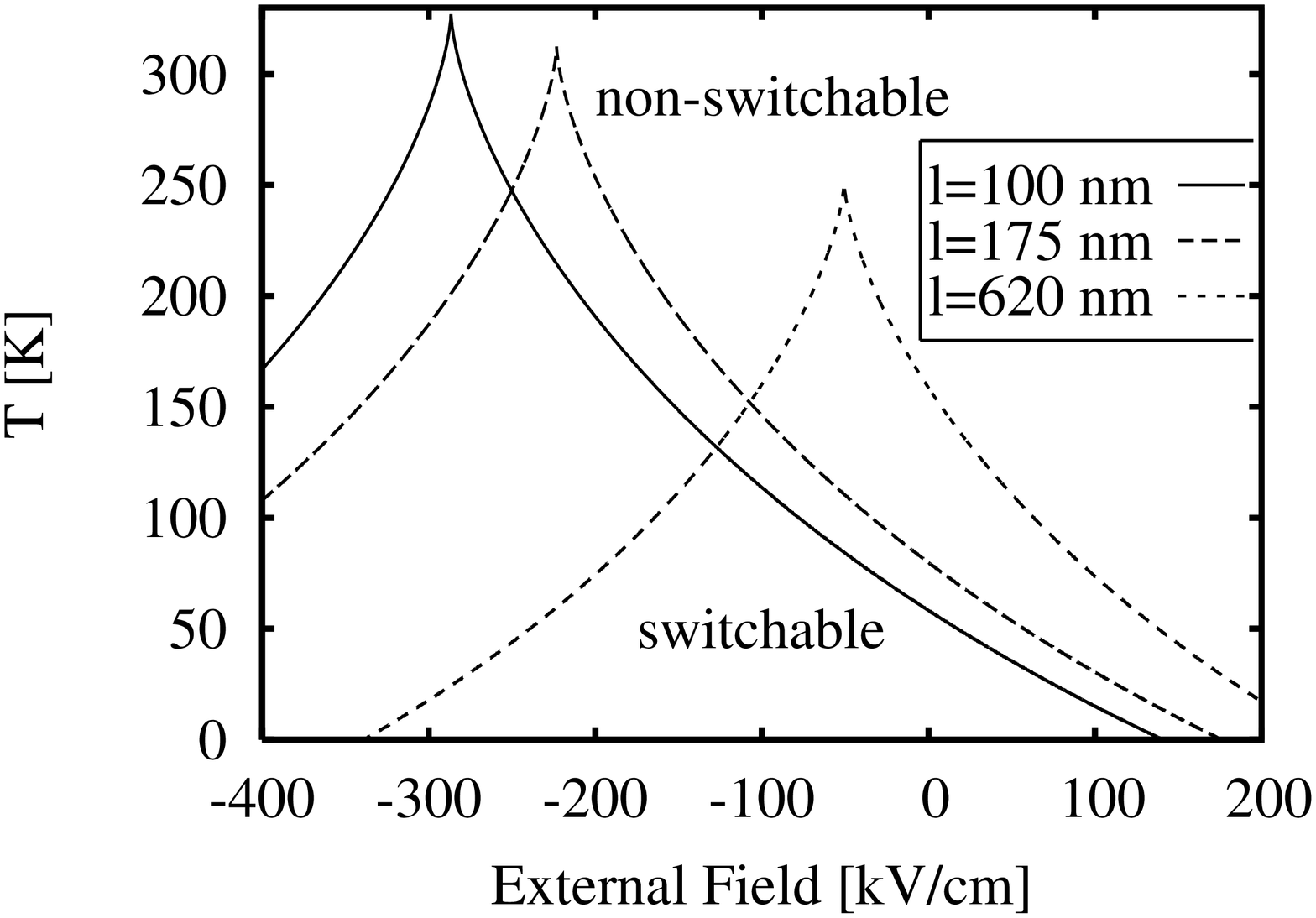}
\caption[tferrod]
        {Transition temperature $T_{ferro}$ as a function of applied external
 electric field $E_{ext}$ for $BST$ films on $SRO$ substrates of thicknesses
 $100$ (solid), $175$
 (dashed) and $620$ nm (dotted line). The corresponding line divides the region into switchable
 and nonswitchable polarization phases for each film.}
\label{tferrodbstfig}
\includegraphics[angle=-90,scale=0.33]{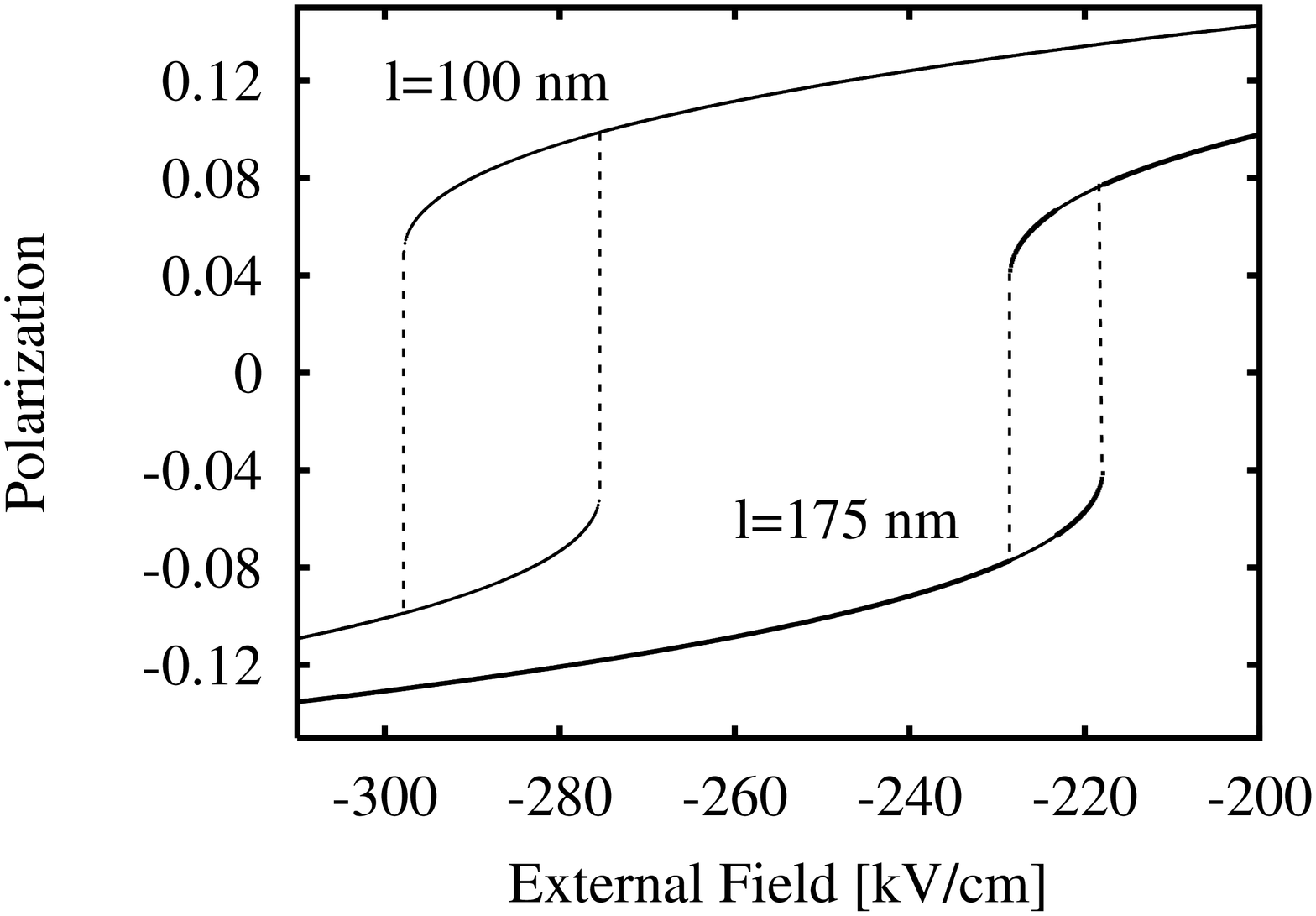}
\caption[pd290]
        {Calculated hysteresis polarization loops for $BST$ films on $SRO$ 
substrates of thicknesses $100$ and $175$ nm at temperature 290 K.}
\label{pd290bstfig}
\end{figure}

In Fig.~\ref{templfig}
we display the three temperature-scales as a function of thickness
for BST films on SRO electrodes with $E_{ext} = 0$:  $T_{max}$, $T_c^*$
and $T_{ferro}$.
Because $E_l^T = W_l \neq 0$, there is a clear separation of
the three temperatures; for $E_l^T = 0$, they collapse onto $T_c^*$
(compare Eqs.~\ref{eqnTc},~\ref{eqnTmax} and~\ref{eqnTferro}).
Therefore an estimate of $W_0$ 
can be obtained from the difference of $T_{max}$ and 
$T_c^*$, where the latter can be
expressed in terms of (experimentally accessible) Landau parameters
and by the value of the misfit strain (Eq.~\ref{eqnTc}).
The magnitude of the obtained bias field, $W_0 = 400$ kV/cm, is in rough 
agreement with experiment: the experimental temperature 
$T_{max}^{exp} = 250$ K 
for a $950$ nm film in Fig.~\ref{pgfig} is close 
to the calculated 
$T_{max}^{calc} = 268$ K in Fig.~\ref{templfig} and
$T_{max}^{exp} = 280$ K 
for the $660$ nm film in Fig.~\ref{pgfig} is also reasonably close to 
the calculated 
$T_{max}^{calc} = 304$ K in Fig.~\ref{templfig}.

The temperature $T_{max}$ displays a peak, $T_{max} = 519$ K at $l^*= 60$ nm, 
and it
decreases with increasing thickness and asymptotically approaches $T_c^*$,
the bulk transition temperature.
As previously noted in Section II, for $l < l^*$, 
$T_{max}$ decreases with decreasing $l$;
by contrast for $l > l^*$, the trend is consistent with that displayed
in Fig.~\ref{pgfig}.
More generally, the behavior of
$T_{max}$ depends on that of the polarization as a function 
of temperature (see Eq.~\ref{eqnTmax}).
Both the strain and the bias field make the polarization
decrease with increasing film thickness ($l > 60$ nm) as will be discussed
shortly, and this monotonic 
behavior makes $T_{max}$ decrease as well.
The observed peak in $T_{max}$ is driven by the depolarization field 
contribution in our model.
$T_{max}$ becomes suppressed for very thin films ($l < 60$ nm) 
and approaches zero just as does the polarization at these film
thicknesses
(see Fig.~\ref{pl300aptbstfig}).
The value of $l^*=60$ nm is determined by the strength of the 
depolarization contribution, and thus is expected to depend on
the electrode/substrate material.
As already noted, we use a longer $l_e$ for BST films on PSS than for BST on SRO.
Because the depolarization field contribution to our free energy
is proportional to $\frac{l_e}{l}$, (see Eq.~\ref{eqnalphad}), 
the resulting $l^*$ will be larger for BST on PSS than on the SRO 
substrate.
For BST on PSS, we calculate that the peak in $T_{max}$ occurs at  
$l^* \sim 300$ nm in contrast to the value of $l^* \sim 60$ nm for BST
on SRO displayed in Fig.~\ref{templfig}.

The theoretical temperature $T_c^*$ in Fig.~\ref{templfig} refers to the 
pa\-ra\-e\-lec\-tric-ferro\-e\-lec\-tric
transition at zero total field, $E_l^T = W_l + E_{ext} = 0$ (see Eq.~\ref{eqnTc}).
It has a peak at $80$ nm with a maximum value of $T_c^* = 329$ K and then 
decreases with increasing film thickness due to the strain relaxation term,
$u_l \sim e^{-l/lc}$.
It reaches its bulk transition temperature value ($T_{bulk} = 235$ K) for thicker
fully relaxed films: $T_c^* = 238$ K for $1000$ nm film.
Due to the depolarization contribution  (see Eq.~\ref{eqnalphad}),
$T_c^*$ is suppressed for very thin films and eventually reaches zero.

The temperature $T_{ferro}$ that separates switchable and 
nonswitchable polarization regimes increases for increasing $l$ 
and saturates 
when it reaches the bulk transition temperature.
In Eq.~\ref{eqnTferro} we see that there is competition 
between thickness-dependent contributions due to
$u_l$
and to $W_l$ in $T_{ferro}$;
therefore
whether this temperature increases or decreases with 
increasing film thickness depends on their relative magnitudes.
For example, $T_{ferro}$ is shown to increase with increasing 
$l$ in Figure~\ref{templfig}, reflecting the dominance of the bias field
contribution in this particular case.
We note that the switchable (ferroelectric) regime is lost as
the films become very thin: e.g. ferroelectricity 
vanishes at $l \sim  100$ nm at fixed temperature $60$ K.
$T_{ferro}$ reaches zero at a critical film thickness 
($l_{crit} \sim 35$ nm here)
and films with $l < l_{crit}$ remain in the nonswitchable 
polarization regime at
all temperatures.

In Fig.~\ref{ertbstfig} we display our calculated relative
permittivity $\epsilon (T)$ 
($\epsilon = 1 + \chi$) for BST films on SRO substrates with $E_{ext}=0$.
Reduction in $\epsilon (l)$ as a function of decreasing
thickness is observed for both theory and experiment 
(Fig.~\ref{pgfig});
$\epsilon_{max}$ decreases and $T_{max}$ increases as $l$ decreases.
Favorable comparison of the calculated
relative permittivity $\epsilon_{max}^{calc}$ can be made to its 
measured analog~\cite{Sinnamon02} $\epsilon_{max}^{exp}$
shown in Fig.~\ref{pgfig}:  
$\epsilon_{max}^{calc}=2035$ at 
$T_{max}=268$ K matches with $\epsilon_{max}^{exp} \approx 1900$ at 
$T_{max}\approx 260$ K for the $950$ nm film, and  
$\epsilon_{max}^{calc}=1044$ at $T_{max}=304$ K matches with the 
$\epsilon_{max}^{exp}\approx 1100$ at $T_{max}\approx 290$ K for the 
$660$ nm film. 

We also predict the thickness-dependent
dielectric behavior of $\epsilon(T)$ for BST films epitaxially grown on 
PSS in Fig.~\ref{ertbstptfig}.
For films of decreasing thickness where $l<l^*$
a systematic reduction in $\epsilon$ is observed and
the peak in the permittivity is shifted to lower temperatures.

Next we present the calculated nonswitchable polarization
in Fig.~\ref{pl300aptbstfig} 
at temperature $300$ K with
$E_{ext}=0$ for  BST films on SRO.
The polarization is roughly proportional to the 
bias field $W_l$ and its value increases with increasing misfit strain
 $u_l$ (see Eq.~\ref{eqnP}). Motivated by experiment, we have modelled
the bias field and misfit strain to decrease with 
increasing film thickness exponentially
($\{W_l,u_l\}\sim e^{-l/l_x}, \, l_x = \{l_w, l_c\}$), and
therefore the polarization also decreases with thicker films.
However, for very thin films
($l < 50$ nm), there is suppression of the polarization ($P$) 
due to depolarization field effects and a peak with a maximum value of 
polarization $P_{max} = 0.2$ [$C/m^2$] at $l = 50$ nm develops.
The temperature-dependence of the polarization for two different film
thicknesses is displayed in 
the inset of Fig.~\ref{pl300aptbstfig}.
These $P(T)$ curves shows good qualitative agreement with analogous 
measurements.~\cite{Catalan04} In general, the polarization decreases with 
increasing temperature.

$T_{ferro}$, the transition temperature
separating the presence of switchable and nonswitchable spontaneous 
polarization, is plotted in
the presence of external electric field for BST films on SRO
in Fig.~\ref{tferrodbstfig}.
$T_{ferro}(E_{ext})$ reaches its maximum at $E_{ext} = -W_l$ (where
$W_l = W_0 e^{-l/l_c}$), 
and decreases symmetrically
about this value in agreement with Eq.~\ref{eqnTferro}; 
we note that the maximum of $T_{ferro}$ decreases for thicker films as
anticipated by the zero-total field ($E_l^T$) results of $T_c^*$ displayed in
Fig.~\ref{templfig}. 
The temperature $T_{ferro}$ at zero external field matches the behavior of 
Fig.~\ref{templfig}. Switchable and nonswitchable polarization phases
 are marked. 

Calculated hysteresis loops are displayed in Fig.~\ref{pd290bstfig}
for BST films on SRO substrates
at $T= 290$ K 
with $l = 100$ nm and 
$l=175$ nm.
According to Fig.~\ref{tferrodbstfig}, 
at this temperature for these film thicknesses, 
the switchable polarization develops only for certain values of nonzero 
external electric field: $E_{ext} = \left\{-298,-275\right\}$ kV/cm for $100$ nm
film and $E_{ext} = \left\{-229,-218\right\}$ kV/cm for $175$ nm film.
The width of the hysteresis loops in Fig.~\ref{pd290bstfig} is given by the 
above field ranges; it decreases with increasing film thickness and shows
good qualitative agreement with experiment.~\cite{Sinnamon02}
Hysteresis loops are symmetric around the point $E_{ext}= -W_l$ and $P=0$,
where the bias field $W_l$ is the thickness-dependent field offset. 
This field offset (specifically, we refer to the shift of the center of the hysteresis loop along the field
axis from zero field position) becomes
larger for thinner films (see Eq.~\ref{eqnWl}) and is temperature independent.
However, the width of the loops shrinks as temperature is increased,
as shown in Fig.~\ref{tferrodbstfig}.
Symmetry in the hysteresis loops around $P=0$ yields zero 
offset in the spontaneous polarization and therefore 
no associated charge offset within the thin film. 
We note that significant charge offsets are observed in 
graded films with designed polarization and strain gradients.~\cite{Ban03}

\subsection{Strained STO}
\label{STOsec}

\begin{figure}[t!]
\includegraphics[angle=-90,scale=0.33]{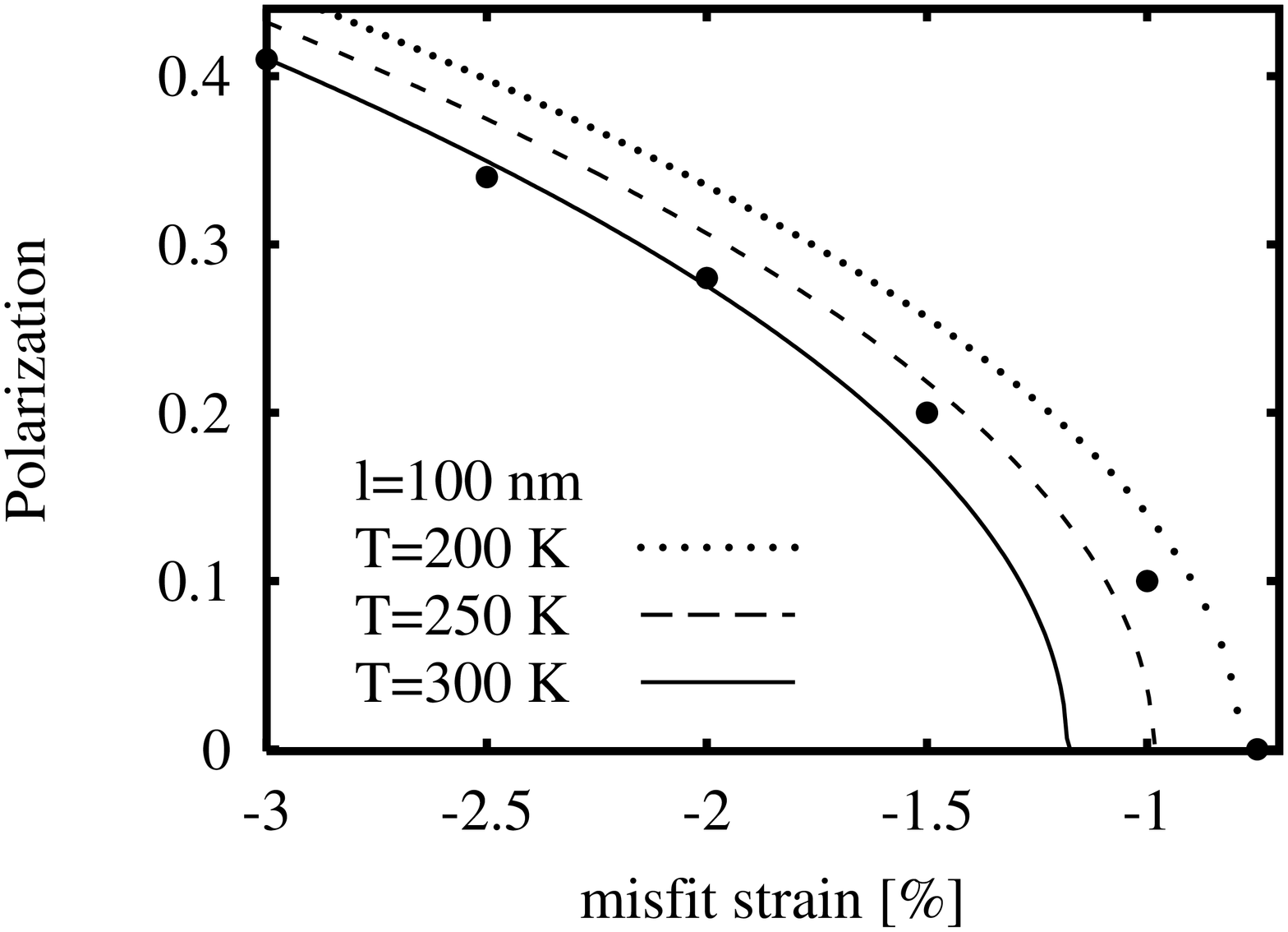}
\caption[pstrainw0sto]
        {Spontaneous polarization $P[C/m^2]$ as a function of film strain 
($u_l$) for $100$ nm STO film at temperatures
$T=200$ (dotted), $T=250$ (dashed)
and $T=300$ K (solid line) and zero total field, $E_l^T=0$. 
Dots correspond to ab-initio values of the polarization~\cite{Antons05}.}
\label{pstrainw0stofig}
\includegraphics[angle=-90,scale=0.33]{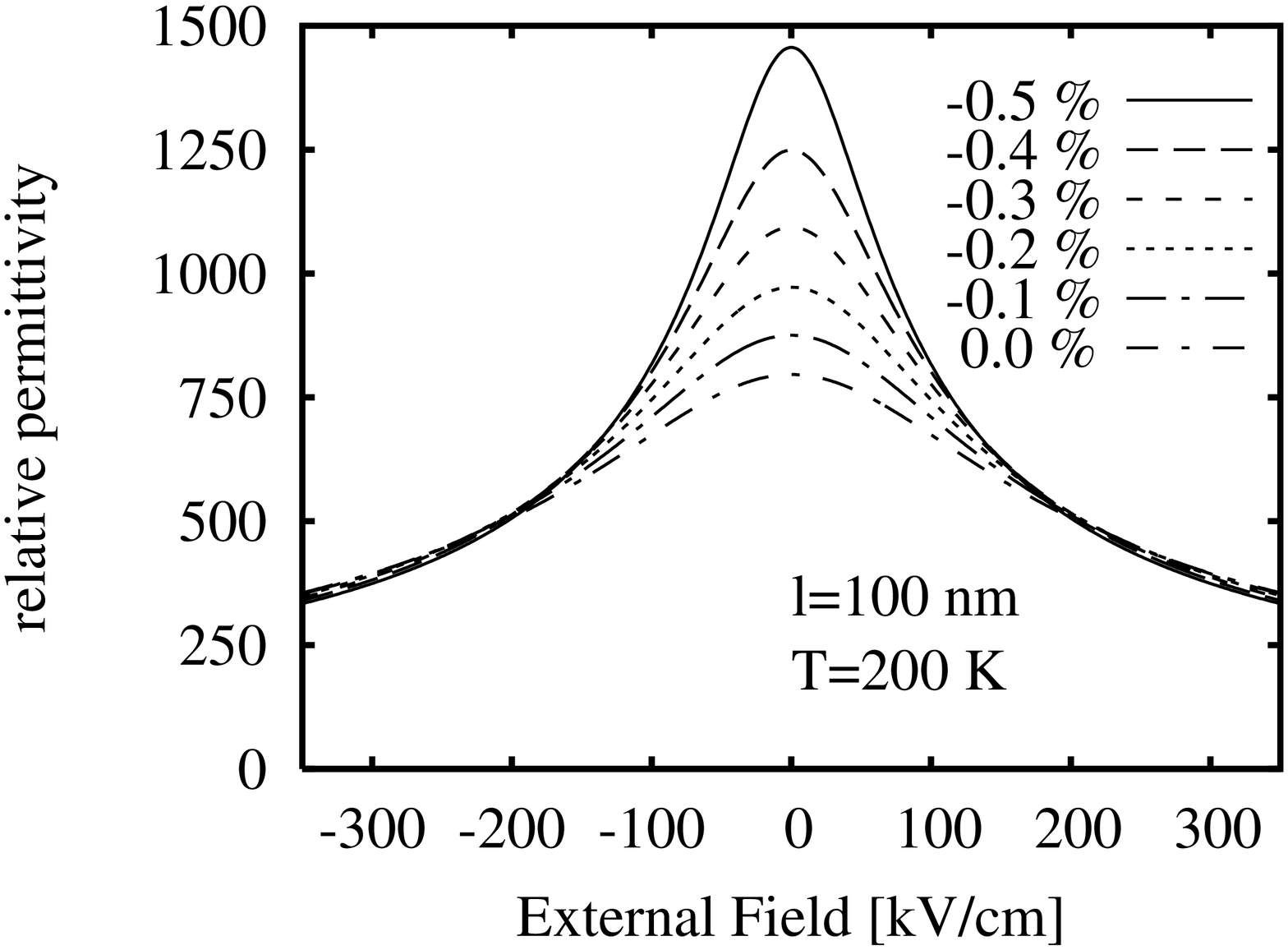}
\caption[erdstrw0para]
        {Calculated paraelectric relative permittivity as a function of 
external electric field $E_{ext}$ and film strain $u_l=-0.5$,
$-0.4$, $-0.3$, $-0.2$, $-0.1$ and $0.0\%$ for $100$ nm STO film at 
temperature $T=200$ K. The permittivity data are shown in the limit of zero 
bias field $W_l$ in order to make comparison with
the ab-initio data~\cite{Antons05}.}
\label{erdstrw0parafig}
\end{figure}

\begin{figure}[t!]
\includegraphics[angle=-90,scale=0.33]{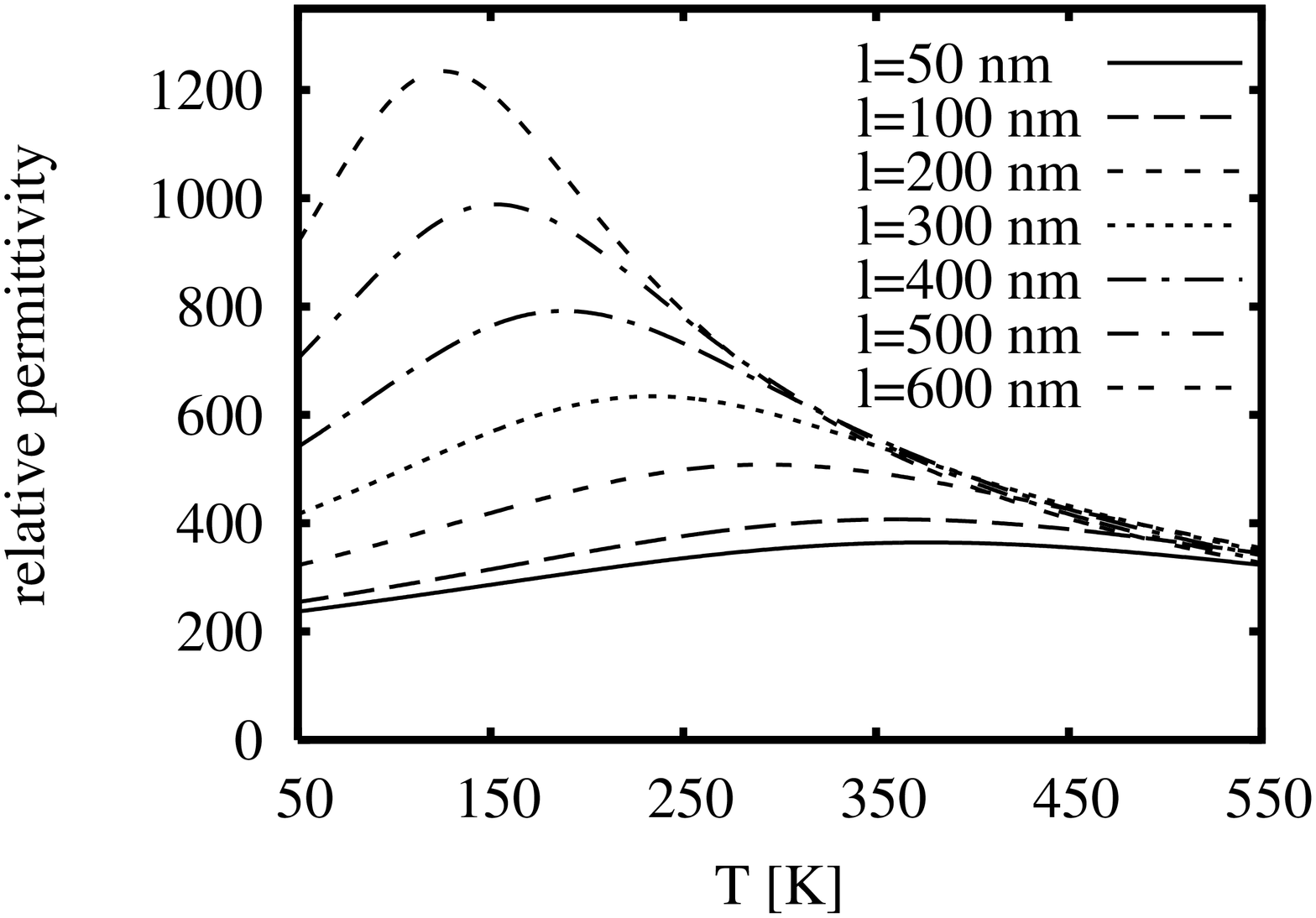}
\caption[ertlsto]
        {Calculated relative permittivity as a function of temperature for
 STO films of thicknesses $l=50$, $100$, $200$, $300$, $400$, $500$ and $600$
 nm at misfit compressive strain $u_m=-0.9\%$ with $E_{ext}=0$. The highest permittivity 
corresponds to the thickest film. Reduction in the permittivity for thin films 
is observed.}
\label{ertlstofig}
\includegraphics[angle=-90,scale=0.33]{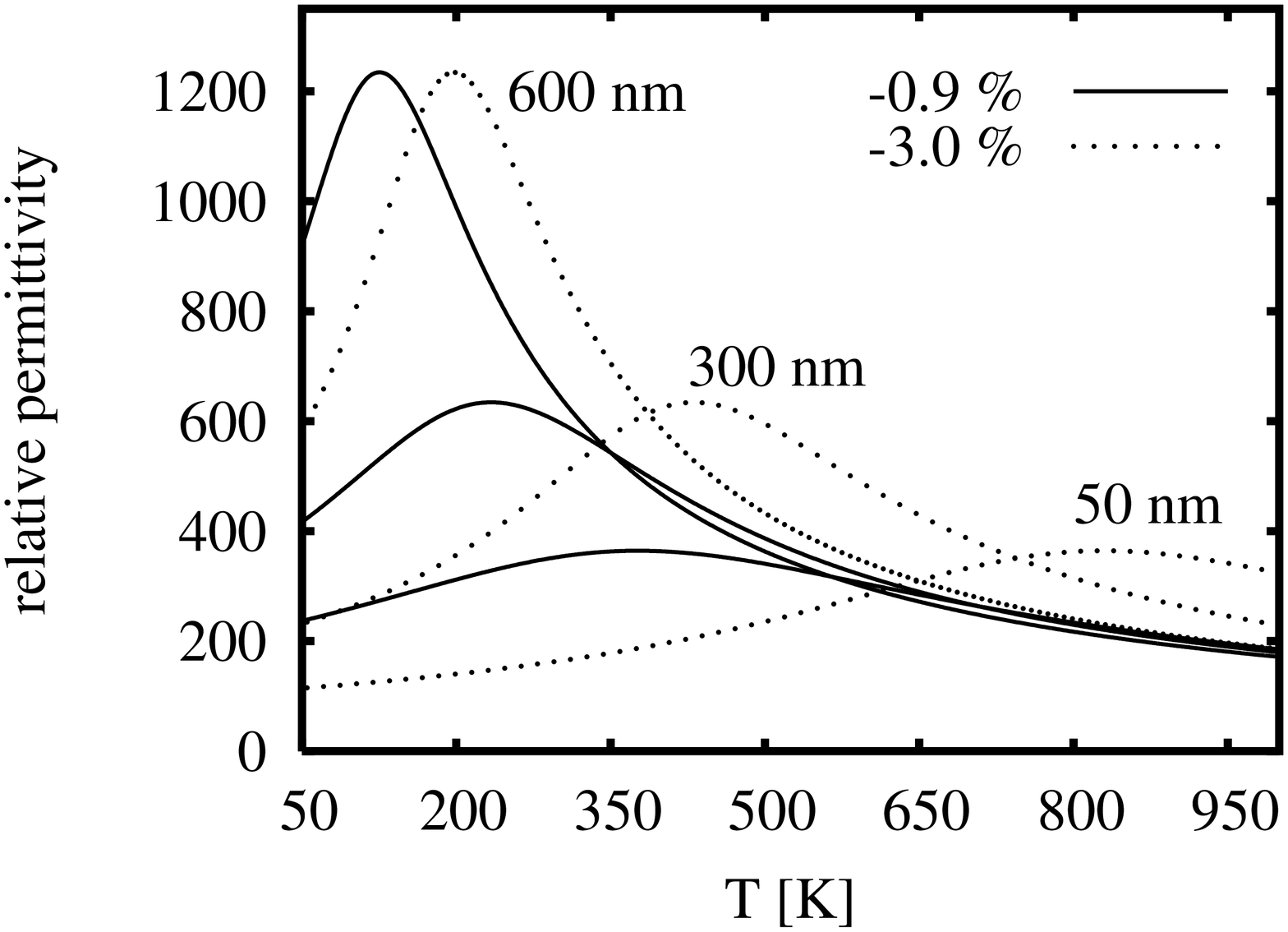}
\caption[ertstrsto]
        {Calculated relative permittivity as a function of temperature for
STO films of thicknesses $l=50$, $300$ and $600$ nm at misfit strain 
$u_m=-0.9\%$ (solid) and $u_m=-3.0\%$ (dotted line) with $E_{ext}=0$. Higher values of misfit
 compressive strain shifts the permittivity curve towards to higher 
temperatures where the larger shifts of $T_{max}$ are observed for the thinner films.}
\label{ertstrstofig}
\end{figure}

\begin{figure}[t!]
\includegraphics[angle=-90,scale=0.33]{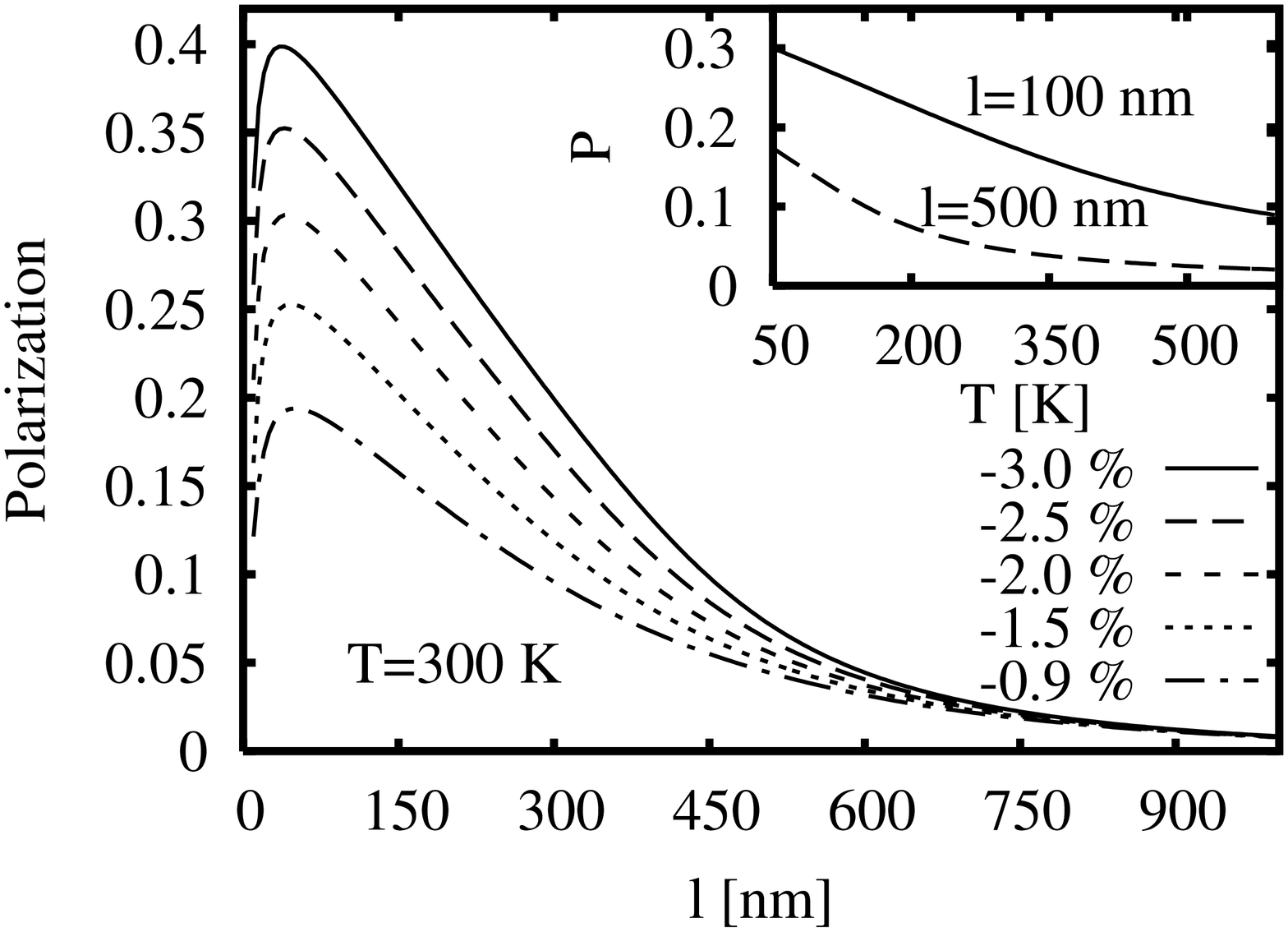}
\caption[pl300tsto]
        {Calculated nonswitchable polarization $P[C/m^2]$ of STO films 
as a function of
film thickness ($l$) for misfit strains $u_m = -3.0$, $-2.5$, $-2.0$, $-1.5$
and $-0.9 \%$ at temperature $300$ K and with $E_{ext}=0$. In the inset,  
the polarization as a function of temperature (T) is plotted for $100$ (solid) and 
$500$ nm (dashed line) films at misfit strain $u_m=-0.9\%$ and zero external 
field.}
\label{pl300tstofig}
\includegraphics[angle=-90,scale=0.33]{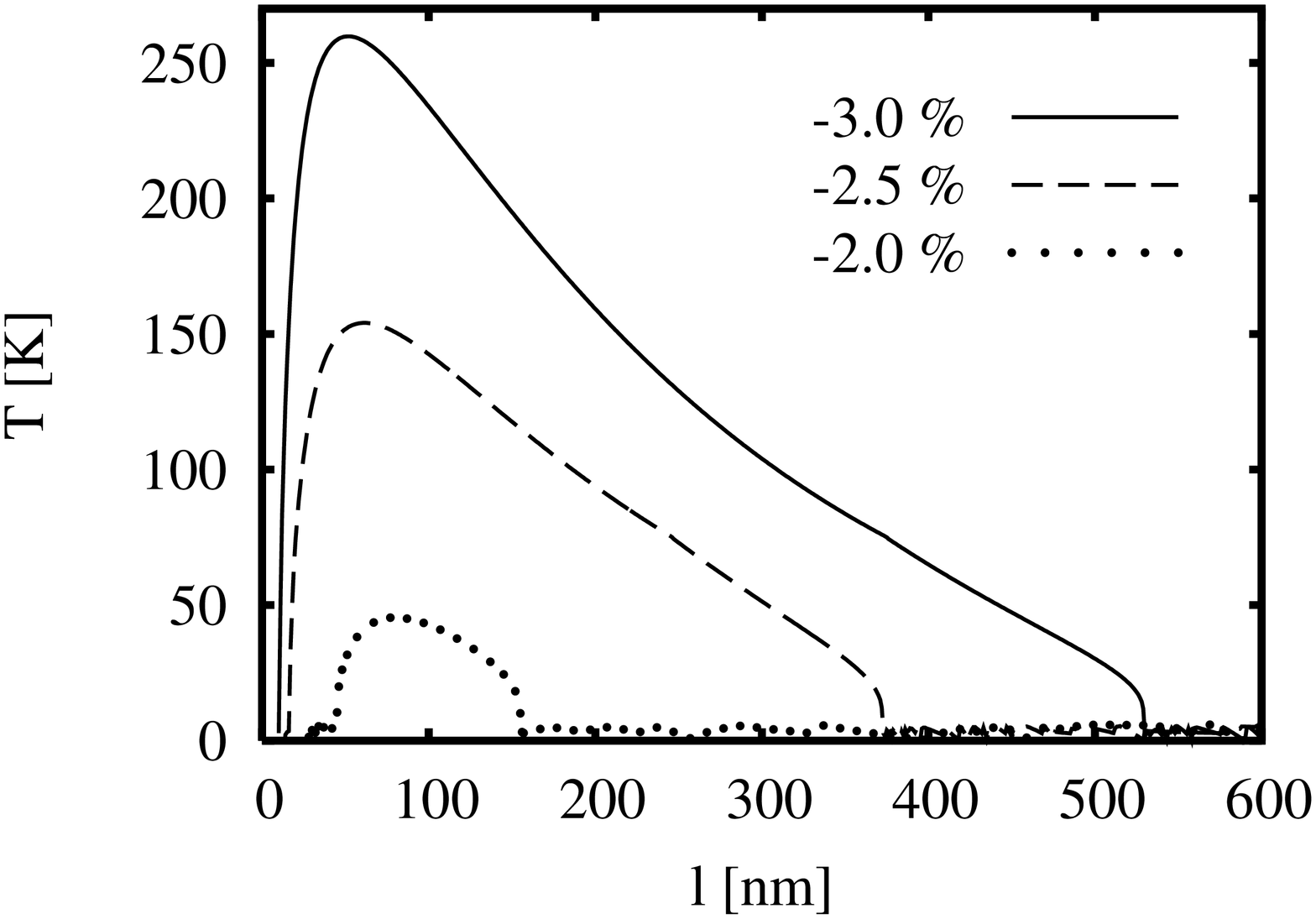}
\caption[tferrolstrainsto]
        {Transition temperature $T_{ferro}$ separating switchable (below) and 
nonswitchable (above the temperature curve)
polarization phases as a function of STO film thickness $l$ and misfit strain
$u_m=-3.0$, $-2.5$ and $-2.0$ $\%$ with $E_{ext}=0$. $T_{ferro}$ becomes 
negative for compressive misfit strain values smaller than $1.5 \%$ for all film 
thicknesses, resulting in a nonswitchable polarization regime.
}
\label{tferrolstrainstofig}
\end{figure}

Bulk strontium titanate $SrTiO_3$ (STO) 
remains
paraelectric down to the lowest temperatures accessible, 
but strained STO films may be driven into a ferroelectric phase even at room 
temperature.~\cite{Haeni04} 
To our knowledge, detailed thickness-dependent
dielectric measurements on such STO films
have been performed only with tensile epitaxial strain resulting to in-plane 
polarization.~\cite{Haeni04,Fuchs99}
Here we make predictions for the thickness-dependent
dielectric properties of STO films 
with compressive epitaxial strains and polarizations normal to
the film-substrate interface.

We begin by making direct comparison between our results and those
of ab-initio studies,~\cite{Antons05}
displayed
in Figs.~\ref{pstrainw0stofig},\ref{erdstrw0parafig}.
Since ab-initio calculations consider uniformly strained films without 
strain relaxation and without an effective bias field, 
we set $W_l = 0$ for the purpose of comparison here.
In Fig.~\ref{pstrainw0stofig} we present the spontaneous polarization
 as a function of misfit strain for a $l = 100$ nm STO film at zero 
total field, $E_l^T=0$. Dots in the figure correspond to first-principles 
calculations,~\cite{Antons05} where the out-of-plane polarization 
in the ferroelectric tetragonal phase ($u_l<-0.75\%$) for  
films with zero macroscopic field has been calculated. 
We choose the $l = 100$ nm STO film 
where we do not expect depolarization effects to be important 
($l > l^*$; see Fig.~\ref{pl300tstofig}) 
for comparison with the ab-initio data.
Good agreement is achieved at temperatures
$T \sim 250$ K; our calculated curves follow the 
behavior of the ab-initio dots. 
At lower temperatures, better agreement is achieved for less 
compressive strain, 
correctly indicating that fully relaxed STO is paraelectric down to
zero temperature.


Continuing our comparison with the results of
ab-initio calculations,~\cite{Antons05}
we display the paraelectric relative permittivity as a function of 
external electric field $E_{ext}$ and film strain $u_l$ for
a $l = 100$ nm STO film at $T = 200$ K in Fig.~\ref{erdstrw0parafig}. 
A nonpolar tetragonal phase develops for strains 
$-0.75\% < u_l < +0.54\%$ according to the ab-initio 
calculations.~\cite{Antons05}
>From Fig.~\ref{pstrainw0stofig}, the best fit for the polarization 
just at the paraelectric-ferroelectric phase boundary ($u_l = -0.75\%$)
is achieved at $T = 200$ K, and therefore
we choose this temperature to calculate our paraelectric permittivity 
data.
We compare our results to the ab-initio calculations 
in the range of compressive strain 
$-0.5\% < u_l < 0.0\%$: 
in both cases, the permittivity ($\epsilon$)
increases with increasing compressive strain;
this occurs even 
more rapidly 
in the range of external fields $-150 < E_{ext} < 150$ kV/cm,
and its magnitude in the range $800 \le \epsilon \le  1500$ in 
Fig.~\ref{erdstrw0parafig} roughly corresponds to 
the ab-initio values $400 \le \epsilon \le 1800$.~\cite{Antons05}
We note that the observed increase of the 
paraelectric permittivity with strain can be understood from 
Eq.~\ref{eqnchi}.

Both Figs.~\ref{pstrainw0stofig} and~\ref{erdstrw0parafig} indicate
good agreement between results of our phenomenological model
and those of previous first-principles calculations, 
and this provides us with confidence regarding the Landau
coefficients and more generally with the approach described 
here applied to
strained STO films.
Next we calculate the dielectric properties of strained 
STO films using a nonzero
value for the effective bias field, $W_0 = 400$ kV/cm, that is 
comparable to that
used for BST films in Sec~\ref{BSTsec}.

Our calculated relative permittivity as a function of temperature and 
film thickness at compressive strain $u_m=-0.9\%$ (STO on LSAT) 
with $E_{ext}=0$ is plotted in Fig.~\ref{ertlstofig}. 
The permittivity is suppressed for thinner films and its maximum is shifted
towards higher temperatures, displaying similar trends 
for both $\epsilon$
and $T_{max}$ as for BST on SRO in 
Sec~\ref{BSTsec}. 
$T_{max}$ increases with decreasing $l$, but develops 
a peak at $l = 60$ nm and is again suppressed for very thin films due to
depolarization effects.
The screening length of LSAT 
is comparable to $l_e$ for SRO, and thus so is $l^*$.
The magnitude of $\epsilon$ is also comparable to that in
BST films in Fig.~\ref{ertbstfig};
it results from similar values of the Landau coefficients and
the value of the compressive strain in both films (see Table~\ref{LGtab}).
To our knowledge, there exists only one published dielectric measurement
on strained STO with the polarization normal
to the film-substrate interface;
this experiment, performed on a  $l = 50$ nm STO film grown on LSAT 
($u_m = -0.9\%$)~\cite{Haeni04}
yields $\epsilon_{max}^{exp}\approx 400$, that is in a good agreement 
with our calculated maximum value of permittivity, 
$\epsilon_{max}^{calc}=364$ for this film/substrate combination.
We investigate the effect of compressive strain on the 
relative permittivity in 
Fig.~\ref{ertstrstofig}. We observe that increased compressive strain 
shifts the permittivity curve towards higher temperatures and larger shifts of
 $T_{max}$ occur for thinner films.

We display the nonswitchable polarization as a function of film thickness and 
misfit strain at $T = 300$ K and $E_{ext} = 0 $ in
Fig.~\ref{pl300tstofig}. Again, the polarization shows similar
behavior as in BST films in Sec.~\ref{BSTsec}; it decreases with 
increasing film thickness and is suppressed for very thin films ($<50$ nm) due
to depolarization effects.
Increasing the strain results in higher polarization, in agreement
with Fig.~\ref{pstrainw0stofig}; this time however the 
polarization values change
due to the effect of a nonzero bias field.
In the inset, the temperature-dependence of the nonswitchable polarization is 
plotted in $l=100$ and $l=500$ nm STO films at misfit strain $u_m=-0.9\%$ 
(STO on LSAT substrate). 
The polarization decreases with increasing temperature. 

Unstrained bulk STO remains paralectric down to zero temperature. However,
as previously noted, a 
ferroelectric regime occurs for strained STO films.~\cite{Haeni04}
We plot the transition temperature $T_{ferro}$, separating switchable and 
nonswitchable polarization regimes, as a function of film thickness and misfit 
strain with $E_{ext}=0$ in Fig.~\ref{tferrolstrainstofig}.
We predict a ferroelectric phase to occur for compressive 
strains larger than $2.0 \%$,
and note that ferroelectricity is recovered here for the thinnest STO films, 
as one goes from a thick-film nonswitchable regime to a thin-film switchable
one (at fixed $T$); e.g. ferroelectricity emerges at $-2.0\%$ 
strained films for thicknesses, $ 30 \lesssim l \lesssim 160$ nm.
This is distinct from the behavior previously described in BST films, 
where ferroelectricity is lost by
making films thinner (see Fig.~\ref{templfig}).
$T_{ferro}$ indicates a maximum at $l^* \approx 60$ nm, and this
peak is due
to depolarization effects (see Eq.~\ref{eqnalphad}) that reduce the
transition temperature to zero for the thinnest films.
$T_{ferro}$ decreases in thicker films ($l>60$ nm) for values of 
compressive strain $u_m > 2.0\%$
as shown in Fig.~\ref{tferrolstrainstofig}. However, it increases 
with increasing film thickness for small values of compressive strain 
($u_m \lesssim 1.5 \%$), similarly to BST films on SRO substrates
 (with $u_m = -0.5\%$) in Fig.~\ref{templfig}.
We note here that although $T_{ferro}$ increases, it has
negative nonphysical value for these low strain values, and thin films remain 
in the nonswitchable regime down to zero temperature.
As noted previously, the thickness-dependence of this temperature
scale arises from competition between strain and bias
field contributions
(see Eq.~\ref{eqnTferro}); the former dominates 
for large enough mistmatch strains ($u_m \gtrsim 2.0\%$), 
and in this case 
$T_{ferro}$ decreases with increasing film thickness.

\section{Discussion}
\label{discussionsec}

\begin{figure} [t!]
\begin{center}
\includegraphics[scale=0.35]{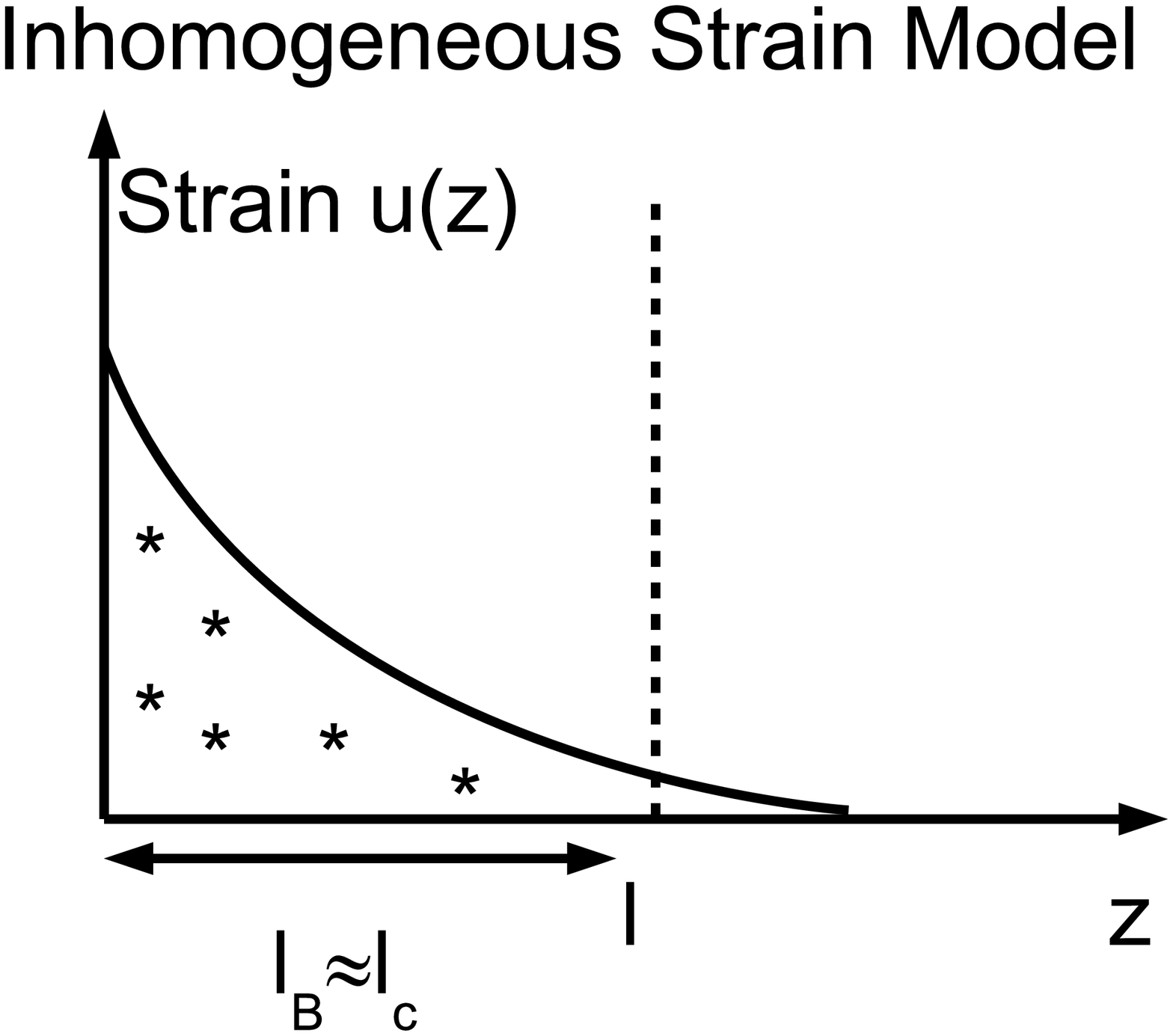}
\caption[catalanstrain2]
{Schematic of the inhomogeneous strain model~\cite{Catalan04}
where
the effective bias field $W(z)$ is spatially-dependent
due to flexoelectric coupling; here the characteristic
length-scales associated with the strain relaxation ($l_c$)
and the buffer layer ($l_B$), where the elastic defects reside,
are assumed comparable with the overall film thickness ($l$).}
\label{catalanstrain2fig}
\end{center}
\end{figure}

Next we explore the implications of our results and the
origin of our model assumptions.
We begin with a general
discussion of the effective bias field in epitaxial
perovskite oxide films.
Both the inhomogeneous (see Fig.~\ref{catalanstrain2fig}) 
and the segregated (see Fig.~\ref{homogstrain2fig})
strain models describe the thickness-dependent dielectric properties
of ferroelectric films consistently with experiment, and thus
further measurements are required to determine the 
presence/absence of underlying inhomogeneous strain 
throughout the film. These models each have effective bias fields,
one that is spatially varying~\cite{Catalan04}
and the other that is uniform,
and it is exactly this feature that we exploit in a proposed benchtop
experiment to distinguish these two scenarios.

An effective bias field breaks up-down symmetry at all 
temperatures. In a film above the zero-(external) field 
($E_{ext} = 0$) ferroelectric transition 
temperature $T_{ferro}$, this results in a nonzero macroscopic polarization 
in zero external electric field (see Eq.~\ref{eqnP}); for the sake
of completeness, we note that the Curie temperature $T_c^*$ refers to
the paraelectric-ferroelectric transition at zero total field 
($E_l^T \equiv W_l + E_{ext} = 0$).  
While this polarization can 
vary with temperature, making the film pyroelectric, it should not be 
confused with a ferroelectric spontaneous polarization. The correct 
distinction between polar and ferroelectric films is made on the 
basis of switchability, as determined, for example, through a hysteresis 
measurement.
In a nonswitchable polar film, there is only one locally stable 
polarization state with $E_{ext} = 0$, and the system will show 
dielectric behavior in a finite external electric field.
On the other hand, if there are two (or more) stable polarization states 
for the system with $E_{ext} = 0$ that can be switched by application 
of an external electric field, then the film is ferroelectric, and it 
will exhibit a characteric ferroelectric hysteresis loop. An effective 
bias field $W_l$ will lead to a lateral shift in the ferroelectric hysteresis 
loop, which can be used to determine the associated effective bias 
voltage across the film. Field offsets have been indeed experimentally 
observed in $100$ nm $PbZr_{0.2}Ti_{0.8}O_3$ (PZT) thin 
films.~\cite{workshop07}
We also remark that an effective bias field will make the two polarization 
states symmetry-inequivalent with different energies. 

The transition from nonswitchable to switchable 
ferroelectric phases usually occurs as a function of
decreasing temperature at 
$T_{ferro}$.
The detailed thickness-dependence of T$_{ferro}$ depends on material-specific 
parameters, as can 
be seen in Figs.~\ref{templfig}
and~\ref{tferrolstrainstofig} for BST and STO thin films. 
More generally the $l$-dependence of the dielectric properties 
enters via the 
strain relaxation
(Eq.~\ref{eqnul}), the bias  (Eq.~\ref{eqnWl}) and 
the depolarization fields.
For the strain relaxation, an exponential decay on a characteristic 
length scale of several
hundred nanometers was observed
experimentally.~\cite{Sinnamon02,Kim99}
In our model, we assume the same exponential decay for the magnitude of 
the uniform effective bias field.
These two quantities determine the thickness-dependence of the quantities 
of interest in all but the very thinnest films, where the  
depolarization field term dominates, strongly suppressing $T_{ferro}$, 
the polarization, and $T_{max}$.
In the case of the temperature $T_{ferro}$
(see Figs.~\ref{templfig} 
and~\ref{tferrolstrainstofig}),
the strain and effective bias contributions opppose each other 
(Eq.~\ref{eqnTferro}) and depending on their
relative strengths, $T_{ferro} (l)$ increases 
(BST case where $W_l$
dominates) or decreases (situation for strained STO
where $u_l$ is greater) with increasing film thickness $l$.

A direct consequence of the 
strain contribution to $T_{ferro}$ is that
we predict that ferroelectricity can 
be strengthened as the films get thinner 
($u_l$ increases with decreasing $l$), 
resulting in a transition from a 
nonswitchable polar phase to a ferroelectric 
state below a critical thickness $l_{CT}$; 
more specifically, for STO measured at 
$100$ K and compressive strain $-2.5\%$ 
(see Fig.~\ref{tferrolstrainstofig}), 
the critical thickness below which ferroelectricity appears 
is $l_{CT} = 200$ nm. 
This runs counter to the usual notion that ferroelectricity is 
suppressed as the film thickness decreases, disappearing below a critical 
thickness; we note that would be the case for BST on SRO
(here $W_l$ dominates the expression for $T_{ferro}$) 
where our results displayed in Fig.~\ref{templfig}
indicate a critical thickness of about $100$ nm at temperature $60$ K.
This reentrant ferroelectricity as a function of decreasing
$l$ should be readily observable in an appropriate experiment
for strained STO films with the polarization normal to the
film-substrate interface.

Both the strain and the effective bias contributions act to
decrease $T_{max}(l)$ and $P_l$ (related by Eq.~\ref{eqnTmax})
as a function of increasing film thickness ($l$) 
(see Figs.~\ref{templfig},~\ref{pl300aptbstfig} and~\ref{pl300tstofig}).
By contrast, the depolarization contribution suppresses 
$T_{max}(l)$ and $P_l$ with decreasing $l$. 
The dominance of this depolarization term explains the observed
shift of the peak of $T_{max}(l)$ to
higher values of $l^*$ 
for semiconducting substrates (e.g. PSS)
(see Fig~\ref{ertbstptfig})
that have longer screening lengths ($l_e$) than their
metallic counterparts (e.g. SRO); here we recall that
the depolarization contribution to the free energy expansion
is $\alpha_d \sim l_e/l$ (Eq.~\ref{eqnalphad}).
We note that the thickness-dependent effect of the depolarization
field on the relative permittivity has been noted before~\cite{Bratkovsky05} 
with a similar term, $\alpha_d^{BL} = a/l$, 
where boundary conditions for the spatially-varied polarization
are proposed that incorporate the effects of a symmetry-breaking 
effective field.
In this previous approach,~\cite{Bratkovsky05}
$a$ then is a boundary-related characteristic length.
Since the thickness-dependence in both treatments is the same, one obtains
similar results for the relative permittivity with appropriate 
choice of these length-scales
($l_e$ and $a$) although their physical origins are different.
Here we have extended this treatment to address the thickness-dependence
of other dielectric properties (e.g. polarization) as well, and we
note that the previous inhomogeneous strain approach~\cite{Catalan04}
did not include such depolarization effects for thin ferroelectric films.

\begin{figure} [t!]
\begin{center}
\includegraphics[scale=0.3]{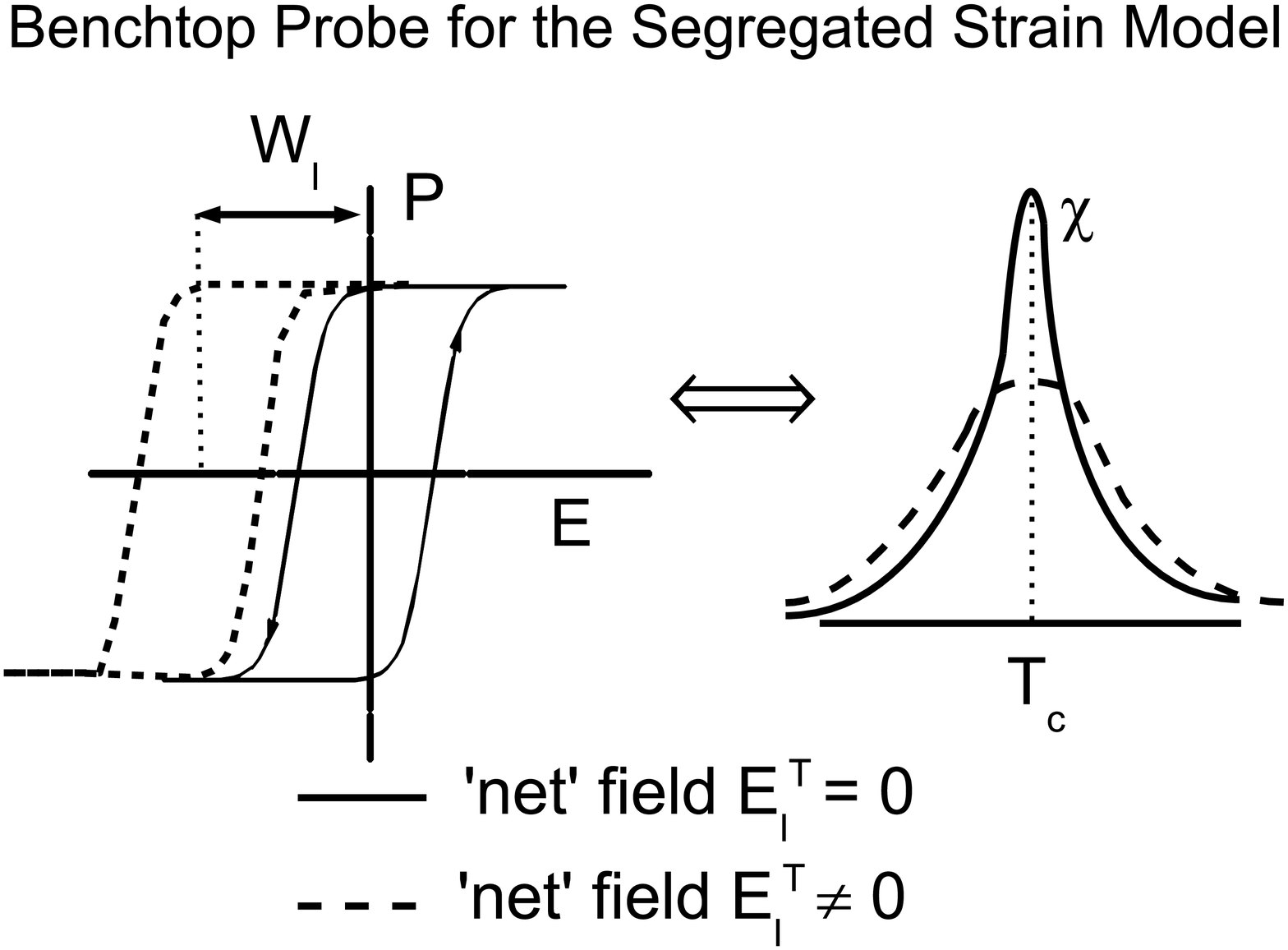}
\caption[hysteresis3]
{Schematic of a benchtop probe to test for
the segregated strain model:
the field offset ($W_l$) in the observed
hysteresis loop can be tuned to zero
by application of an electric field $E^*_{ext}(l) = -W_l$;
in this case, the relative permittivity sharpens
since the net 
(thickness-dependent) field $E_l^T = E^*_{ext}(l) + W_l = 0$.}
\label{hysteresis3fig}
\end{center}
\end{figure}

The smearing of the sharp peak in the temperature-dependent dielectric 
response(Figs.~\ref{ertbstfig},~\ref{ertbstptfig} and~\ref{ertlstofig})
in zero electrical field ($E_{ext} = 0$) is a signature of the presence
of a finite effective bias field $W$; this point has been much
discussed
previously both here (see Eqs.~\ref{eqnchi},~\ref{eqnP}) and 
by others.~\cite{Catalan04,Bratkovsky05,Vendik00,Saad04}
$W_l$ assumes larger values for thinner films (see 
Eq.~\ref{eqnWl}) and so pushes the permittivity to smaller values 
in thinner films (see 
Eq.~\ref{eqnchi}), in accordance with experiment (Fig.~\ref{pgfig}).
In the bulk limit, both the strain and the bias field
vanish and bulk behavior of the dielectric properties is recovered.

While there is general agreement that effective bias fields play an 
important role in the properties of 
perovskite thin films,~\cite{Catalan04,Bratkovsky05,Vendik00,Saad04}
their specific origins and their spatial natures in the films are less well 
understood. For example, an effective bias field can be
produced by
a spatially (z)-dependent strain via a 
flexoelectric effect;~\cite{Catalan04} 
we will refer to this as the inhomogeneous strain 
model and it is schematically depicted in Fig.~\ref{catalanstrain2fig}.
In this scenario, misfit dislocations are distributed 
roughly isotropically throughout the film and produce
strain gradients.  
By contrast, in the segregated strain model presented here,
the elastic defects are concentrated in a thin buffer layer near
the film-substrate interface 
(see Fig.~\ref{filmfig} and 
Fig.~\ref{homogstrain2fig}).
This buffer layer itself breaks the up-down symmetry of the film, 
which then results in a nonzero effective field.
To be more specific, it may be that an edge dislocation in the buffer 
layer produces a local polar distortion. This makes the buffer layer 
polar and produces a field in the uniform film. We then expect the strength
of the effective
bias field to be related to the areal density of dislocations, 
and thus to the magnitude of the homogeneous
strain in the film.

As we have shown in Sec.~\ref{BSTsec}, using the segregated
strain model (see Figs~\ref{filmfig} and~\ref{homogstrain2fig})
we recover the 
thickness-dependent dielectric
properties of BST films consistent with those
measured~\cite{Sinnamon02} and calculated using flexoelectric 
effects~\cite{Catalan04} within an inhomogeneous strain scenario 
(see Fig~\ref{catalanstrain2fig}).
Therefore, in order to determine which of these two models 
is applicable to a particular 
film, additional experimental characterization is necessary. The 
presence of strain gradients in the BST films 
studied using flexoelectricity~\cite{Catalan04} was verified by x-ray  
analyses~\cite{Catalan05}, though similar studies on
different BST films (on different substrates) indicate the absence of
such inhomogeneous strain.~\cite{Balzar02}  Therefore it is clear
that observations of thickness-dependent permittivities and polarizations
are not enough to indicate the underlying strain profile of
the ferroelectric film.  Naturally the presence/absence of
strain gradients can be addressed directly by x-ray
diffraction but this probe may not be always easily 
accessible or practical.

Here we emphasize that the effective bias fields associated
with these two strain models are spatially different
(see Figs~\ref{catalanstrain2fig} 
and~\ref{homogstrain2fig}),
and we will use this distinction to propose a benchtop
experiment to distinguish between these two scenarios.
In the inhomogeneous strain model,~\cite{Catalan04} 
the effective bias 
field is spatially nonuniform; more specifically
is proportional to the strain gradient via a flexoelectric
coupling, and thus has an exponentially decaying spatial
profile within the film (see Fig~\ref{catalanstrain2fig}).
By contrast, in the 
segregated strain model (see Fig~\ref{homogstrain2fig}), 
the effective bias field is uniform from the 
edge of the buffer layer to the surface. 
Assuming that this effective bias field $W_l$ is uniform in
the majority of the film, one can tune an external electric
field to the right value, $E^*_{ext} (l) = - W_l$, to create
a net zero-field condition ($E_T (l)  = E^*_{ext} (l) + W_l$)
where the specific value of the necessary external field
would be thickness-dependent.
Therefore there would exist an external field value $E^*_{ext} (l)$
when the hysteresis loop would no longer have a field offset;
at this value of the applied external field, a sharp peak
in the temperature-dependence of the relative permittivity
should be observed (Fig~\ref{hysteresis3fig}).
We emphasize that this must be a unipolar hysteresis experiment,
with single top and bottom electrodes; many measurements are
performed across two top electrodes, namely two series capacitors,
where one would always be uncompensated.
Another probe of the spatial uniformity 
of $W$ would be to measure T$_{ferro}$ as a function 
of $E_{ext}$; for a uniform effective bias field there 
would be a sharp peak in T$_{ferro}$, as shown in 
Fig.~\ref{tferrodbstfig}.
This pronounced peak would not be present 
for a spatially-varying effective bias field $W(z)$
since the latter would have varying magnitude in the sample and no 
particular value
of applied (uniform) $E_{ext}$ could completely compensate for it
everywhere in the film; physically we note that $W(z)$ 
could arise from coupling of the polarization to gradients
in the strain,~\cite{Catalan04} in chemical composition,
and in temperature.~\cite{Bratkovsky05,Ban03,Alpay03}  
There do exist graded ferroelectric structures where such spatially
varying quantities
are explicitly present by design; here a charge offset is
often observed in the charge-voltage 
hysteresis loops,~\cite{Ban03} and 
this could serve as an indicator of underlying gradients
in ferroelectric films if a suitable ``charge origin'' could
be chosen as a reference.

We emphasize that we expect different films, with varying 
compositions, substrates and growth conditions, 
to have diverse strain and effective bias field profiles. 
The inhomogeneous scenario may describe some while 
others may be better modelled by the 
segregated strain approach; still others may exhibit intermediate 
behavior. By carefully monitoring
growth conditions, it should be possible to control the density and
spatial distribution of 
strain-relieving defects; in some case, it may even be feasible
to kinetically inhibit them to 
obtain uniform coherently-strained films.~\cite{Haeni04}
We emphasize that in each case, the strain and effective bias field 
distributions must be carefully characterized for a full interpretation 
of the measured thickness-dependent dielectric behavior, and
we have presented simple proposals for benchtop measurements
to ascertain the importance of strain gradients in the films.

\section{Summary}

In conclusion,  
we have demonstrated that a segregated strain
model (cf. Fig~\ref{filmfig})
describes the observed thickness-dependent dielectric properties
of ferroelectric films as well as does a previous model
of inhomogeneous strain.~\cite{Catalan04}  
Therefore such thickness-dependent
behavior is not signatory of underlying strain gradients, and
more measurements must be performed to determine the 
strain profile in the film. If the effective bias field is spatially
uniform, it can be compensated by the application of an applied external
electric field $E^*_{ext} (l)  = -W_l$ whose value will be 
dependent on the overall film thickness $l$.  
Benchtop experiments performed 
with $E_{ext} = E_{ext}^*(l)$ will
yield bulk-like sharp dielectric responses.  However such compensation
will not be possible if the effective bias field is spatially varying,
since then its effects cannot be cancelled by the application of an
external
uniform field.  

We have compared our results with experiment (BST on SRO)
whenever possible
and have also made predictions for measurements on strained
STO films with out-of-plane polarization.  Agreement with existing
ab-initio calculations, when appropriate, has been good.  
The possibility of reentrant ferreoelectricity in strained
STO films has also been discussed and we hope that this
will be explored experimentally in the near future.

Our phenomenological study of planar films 
suggests that their thickness-dependent 
dielectric properties are not indicative of 
underlying inhomogeneous
strain, and are consistent with other strain 
profiles.  We view this project as the beginning 
of a broader study of the physical
consequences of boundary-induced effects
in ferroelectrics of increasingly complex
host topologies. A next step is to explore 
cases where the
strain gradients will be induced by geometry:
examples include curved films and cylindrical shells.
Because of the coupling between the elastic
and the electrical degrees of freedom in these
systems, we expect tunable strain gradients
to stabilize novel polarization configurations
with rich phase behavior, and here flexoelectric
effects should definitely be investigated.
More complex host geometries and boundary conditions
are expected to favor more novel orderings and 
dielectric properties; the possibility
of identifying and characterizing these features in
three-dimensional ferroelectrics on the nanoscale
could also be useful in the design of future ferroelectric
memories.~\cite{Arimoto04} 

We thank G. Catalan, M. Dawber, D. Hamman, J. Junquera, V. Kiryukin, 
M. P\'al, J. Scott, and D. Vanderbilt
for discussions.  
We are grateful to I2CAM for support (LP) and to the Aspen Center for Physics
for hospitality (PC, KMR).
We also acknowledge support from grants NSF-DMR-0645461 (LP), 
NSF-NIRT-ECS-0608842 (PC) and 
NSF-DMR-0507146 (KMR).


\begin{thebibliography}{10}

\bibitem{Auciello98} O. Auciello, J.F. Scott and R. Ramesh, {\sl Physics Today}
\textbf{51}, 22 (1998).

\bibitem{Scott00} J.F. Scott, {\sl Ferroelectric Memories} (Springer, Berlin, 2000).

\bibitem{Lines77} M.E. Lines and A.M. Glass, {\sl Principles and Applications of Ferroelectrics and Related Materials} (Clarendon Press, Oxford, 1977).

\bibitem{Pertsev98} N.A. Pertsev, A.K. Tagantsev and N. Setter, 
{\sl Phys. Rev. B} \textbf{80}, 1988 (1998).

\bibitem{Choi04} K.J. Choi, M. Biegalski, Y.L. Li, A. Sharan, J. Schubert, R. Uecker, P. Reiche, Y.B. Chen, X.Q. Pan, V. Gopalan, L.-Q. Chen, D.G. Schlom and C.B. Eom, {\sl Science} \textbf{306}, 1005 (2004).

\bibitem{Haeni04} J.H. Haeni, P. Irvin, W. Chang, R. Uecker, P. Reiche, 
Y.L. Li, S. Choudhury, W. Tian, M.E. Hawley, B. Craigo, A.K. Tagantsev, X.Q. Pan, S.K. Streiffer, L.Q. Chen, S.W. Kirchoefer, J. Levy and D.G. Schlom,  {\sl Nature} \textbf{430}, 758 (2004).

\bibitem{Balzar04} D. Balzar, P.A. Ramakrishnan and A.M. Hermann, {\sl Phys. Rev. B} \textbf{70}, 092103 (2004).

\bibitem{Alpay04} S.P. Alpay, I.B. Misirlioglu, V. Nagarajan and R. Ramesh, {\sl Appl. Phys. Lett.} {\textbf 85} 
2044 (2004); V. Nagarajan, C.L. Jia, H. Kohlstedt, R. Waser, I.B. Misirlioglu, S.P. Alpay and R. Ramesh, {\sl Appl. Phys. Lett.} {\textbf 86} 192910 (2005).

\bibitem{Catalan04} G. Catalan, L.J. Sinnamon and J.M. Gregg, {\sl  J. Phys. 
Condens. Mat.} \textbf{16}, 2253 (2004); G. Catalan, B. Noheda, J. McAneney, L.J. Sinnamon and J.M. Gregg, {Phys. Rev. B} 
\textbf{72}, 020102 (2005).

\bibitem{Bratkovsky05} A.M. Bratkovsky and A.P. Levanyuk, {\sl Phys. Rev. Lett.} {\textbf 94}, 107601 (2005).

\bibitem{Dawber05} M. Dawber, K.M. Rabe and J.F. Scott, {\sl Rev. Mod. Phys.} \textbf{77}, 1083 (2005).

\bibitem{Suzuki99} T. Suzuki, Y. Nishi and M. Fujimoto {\sl Philos. Mag. A} {\bf 79} 2461 (1999).

\bibitem{Sun04} H.P. Sun, W. Tian, X.Q. Pan, J.H. Hanei and D.G. Schlom, {\sl Appl. Phys. Lett.} {\bf 84} 3298 (2004).

\bibitem{Nagarajan05} V. Nagarajan, C.L. Jia, H. Kohlstedt, R. Waser, I.B. Misirlioglu and S.P. Alpay, {\sl Appl. Phys. Lett.} {\textbf 86} 192910 (2005).


\bibitem{Ma01} W. Ma and L.E. Cross, {\sl Appl. Phys. Lett.} {\textbf 78}, 2920 (2001); {\sl ibid} {\textbf 79}, 4420 (2001); {\sl ibid} {\textbf 81} 3440 (2002); {\sl ibid} {\textbf 82} 3293 (2003).

\bibitem{Gruverman03} A. Gruverman, B.J. Rodriguez, A.I. Kingdon, R.J. Nemanich, A.K. Tagantsev, J.S. Cross and M. Tsukada, {\sl Appl. Phys. Lett.} 
{\textbf 83}, 728 (2003).

\bibitem{Tagantsev85} A.K. Tagantsev, {\sl Sov. Phys. JETP} {\textbf 61} 1246 (1985); {\sl Phys. Rev. B} {\textbf 34} 5883 (1986).

\bibitem{Speck94} J.S. Speck and W. Pompe, {\sl J. Appl. Phys.}{\textbf 76} 466 (1994).

\bibitem{Vendik00} O. Vendik and S.P. Zubko {\sl J. Appl. Phys.} {\bf 88} 5343 (2000).

\bibitem{Ahn97} C.H. Ahn, T. Tybell, L. Antognazza, K. Char, R.H. Hammond,
M.R. Beasley, O. Fischer and J.-M. Triscone, {\sl Science} {\bf 276}, 
1100 (1997).

\bibitem{Shaw99} T.M. Shaw, Z. Suo, M. Huang, E. Liniger, R.B. Laibowitz
and J.D. Baniecki, {\sl Appl. Phys. Lett} {\bf 75} 2129 (1999).

\bibitem{Paruch01} P. Paruch, T. Tybell and J.-M. Triscone, {\sl Appl. Phys. Lett.} {\bf 79} 530 (2001).


\bibitem{Sinnamon02} L.J. Sinnamon, R.M. Bowman and J.M. Gregg, 
{Appl. Phys. Lett.} {\bf 81}, 889 (2002).

\bibitem{Lookman04} A. Lookman, R.M. Bowman, J.M. Gregg, J. Kut, S. Rios, M. Dawber, A. Ruediger and J.F. Scott, {\sl J. Appl. Phys.} 
{\bf 96} 555 (2004).

\bibitem{Saad04} M.M. Saad, P. Baxter, R.M. Bowman, J.M. Gregg, 
F.D. Morrison and J.F. Scott, {\sl J. Phys:  Cond. Mat.} {\bf 16}, L451 
(2004).

\bibitem{Chandra06} P. Chandra and P.B. Littlewood, cond-mat/0609347, to
be published in {\bf Physics of
Ferroelectrics:  A Modern Perspective}, eds. K.M. Rabe, C. H. Ahn and J-M. Triscone (Springer, Berlin 2007).

\bibitem{Kim99} H.J. Kim, S. Oh and H.M. Jang, {\sl Appl. Phys. Lett.} {\bf 75} 3195 (1999).

\bibitem{Johnston05} K. Johnston, X. Huang, J.B. Neaton and K.M. Rabe, {\sl Phys. Rev. B} {\bf 71} 100103 (2005).

\bibitem{Rios03} S. Rios, A. Ruedinger, A.Q. Jiang, J.F. Scott, H. Lu and Z. Chen, {\sl J. Phys. Cond. Mat.} {\bf 15} L305 (2003).

\bibitem{Mehta73} R.R. Mehta, B. Silverman and J.T. Jacobs, {\sl J. Appl. Phys.} {\bf 44} 3379 (1973). 

\bibitem{Kretchmer79} R. Kretschmer and K. Binder {\sl Phys. Rev. B} 
{\bf 20} 1065 (1979). 

\bibitem{Dawber03} M. Dawber, P. Chandra, P.B. Littlewood and J.F. Scott, 
{\sl J. Phys. Cond. Mat.} {\bf 15} {L393} (2003); P. Chandra, M. Dawber, 
P.B. Littlewood and J.F. Scott, {\sl Ferroelectrics} {\bf 313}, 7 (2004).

\bibitem{Junquera03} J. Junquera and Ph. Ghosez, {\sl Nature} {\bf 422} 506 (2003).

\bibitem{Tilley92} D.L. Tilley, {\sl Ferroelectrics} {\bf 134}, 313 (1992);
D.L. Tilley in {\sl Ferroelectric Ceramics}, eds. N. Setter and E.L. Colla (Brikhauser Verlay, Basel 1993) pp. 163-184.

\bibitem{Antons05} A.Antons, J.B. Neaton, K. M. Rabe, and D. Vanderbilt {\sl Phys. Rev. B} {\bf 71} 024102 (2005).

\bibitem{Lichtensteiger05} C. Lichtensteiger, J.-M. Triscone, J. Junquera and P. Ghosez, {\sl Phys. Rev. Lett.} {\bf 94} 047603 (2005).

\bibitem{Chen06} L.Q. Chen in 
{\bf Physics of
Ferroelectrics:  A Modern Perspective}, eds. C. H. Ahn, K.M. Rabe and J-M. Triscone (Springer, Berlin 2007).

\bibitem{Fuchs99} D. Fuchs, C.W. Schneider, R. Schneider, and H. Rietschel {\sl J. Appl. Phys.} {\bf 85} 7362 (1999).

\bibitem{workshop07} H.N. Lee, S. Nakhmanson, M.F. Chisholm, 
H.M. Christen, K.M. Rabe and D. Vanderbilt, {\sl Fundamental Physics of Ferroelectrics} (2007).

\bibitem{Catalan05} G. Catalan, B. Noheda, J.McAneney, L.J. Sinnamon and J.M. Gregg, {\sl Phys. Rev. B} {\bf 72} 020102 (2005).

\bibitem{Balzar02} D. Balzar, P.A. Ramakrishnan, P. Spagnol, S. Mani, 
A.M. Hermann and M.A. Matin, {\sl Jpn. J. Appl. Phys.} {\bf 41} 6628 (2002).

\bibitem{Ban03} Z-G. Ban, S.P. Alpay and J.V. Mantese, {\sl Phys. Rev. B} {\bf 67}, 184104 (2003);
Z.-G. Ban, S.P. Alpay and J.V. Mantese, {\sl Integ. Ferr.} 
{\bf 58} 1281 (2003).

\bibitem{Alpay03} S.P. Alpay, Z.-G. Ban and J.V. Mantese, {\sl Appl. Phys. Lett.} {\bf 82} 1269 (2003).

\bibitem{Arimoto04} Y. Arimoto and H. Ishiwara, {\sl MRS Bulletin} {\bf 29} 823 (2004).
 
\end{thebibliography}
\end{document}